\documentclass[journal]{IEEEtran}
\usepackage[colorlinks,linkcolor=blue,anchorcolor=blue,citecolor=blue,bookmarks=true]{hyperref}  
\usepackage[T1]{fontenc}
\ifCLASSINFOpdf
\usepackage{caption }
\usepackage{fancyhdr}
\usepackage[pdftex]{graphicx} 
\DeclareGraphicsExtensions{.eps,.pdf,.jpeg,.png}
\else
\usepackage[dvips]{graphicx}
\fi
\usepackage[compress,nospace]{cite}
\usepackage[cmex10]{amsmath}
\usepackage{epstopdf,amsthm,stfloats,siunitx,amssymb,wasysym,algorithm,algorithmic,array,url, color,subfigure} 
\usepackage{footnote}
\usepackage{booktabs}
\usepackage{balance}
\makesavenoteenv{tabular}
\usepackage{soul}
\newtheorem{myDef}{Definition}
\newtheorem{myProposition}{Proposition}

\begin{document}
\title{NoncovANM: Gridless DOA Estimation for LPDF~System}

\author{Yangying Zhao,~\IEEEmembership{Student Member,~IEEE}, Peng Chen,~\IEEEmembership{Senior Member,~IEEE}, Zhenxin Cao,~\IEEEmembership{Member,~IEEE}, Xianbin~Wang,~\IEEEmembership{Fellow,~IEEE}

\thanks{This work was supported in part by the Natural Science Foundation for Excellent Young Scholars of Jiangsu Province (Grant No. BK20220128), the Open Fund of ISN State Key Lab (Grant No. ISN24-04), the Industry-University-Research Cooperation Foundation of The Eighth Research Institute of China Aerospace Science and Technology Corporation (Grant No. SAST2021-039), the National Key R\&D Program of China (Grant No. 2019YFE0120700), and the National Natural Science Foundation of China (Grant No. 61801112). (\textit{Corresponding author: Peng Chen})}
\thanks{Yangying~Zhao and Peng~Chen are with the State Key Laboratory of Millimeter Waves, Southeast University, Nanjing 210096, China, and also with State Key Laboratory of Integrated Services Networks, Xidian University, Xi'an 710071, China (email: \{zhao\_yangying, chenpengseu\}@seu.edu.cn).}
\thanks{Zhenxin~Cao is with the State Key Laboratory of Millimeter Waves, Southeast University, Nanjing 210096, China (email: caozx@seu.edu.cn).}
\thanks{Xianbin~Wang is with the Department of Electrical and Computer Engineering, Western University, London, ON N6A 3K7, Canada (email: xianbin.wang@uwo.ca).}
}

\markboth{IEEE Transactions on Vehicular Technology}%
{Shell \MakeLowercase{\textit{et al.}}: A Sample Article Using IEEEtran.cls for IEEE Journals}

\maketitle
\begin{figure*}[b]
	\footnoterule
	\footnotesize\textsuperscript{Copyright (c) 2015 IEEE. Personal use of this material is permitted. However, permission to use this material for any other purposes must be obtained from the IEEE by sending a request to pubs-permissions@ieee.org.}
\end{figure*}

\begin{abstract}
Direction of arrival (DOA) estimation is an important research in the area of array signal processing, and has been studied for decades. High resolution DOA estimation requires large array aperture, which leads to the increase of hardware cost. Besides, high accuracy DOA estimation methods usually have high computational complexity. In this paper, the problem of decreasing the hardware cost and algorithm  complexity is addressed. First, considering the ability of flexible controlling the electromagnetic waves and low-cost, an intelligent reconfigurable surface (IRS)-aided low-cost passive direction finding (LPDF) system is developed, where only one fully functional receiving channel is adopted. Then, the sparsity of targets direction in the spatial domain is exploited by formulating an atomic norm minimization (ANM) problem to estimate the DOA. Traditionally, solving ANM problrm is complex and cannot be realized efficiently. Hence, a novel nonconvex-based ANM (NC-ANM) method is proposed by gradient threshold iteration, where a perturbation is introduced to avoid falling into saddle points. The theoretical analysis for the convergence of the NC-ANM method is also given. Moreover, the corresponding Cram\'{e}r-Rao lower bound (CRLB) in the LPDF system is derived, and taken as the referred bound of the DOA estimation. Simulation results show that the proposed method outperforms the compared methods in the DOA estimation with lower computational complexity in the LPDF system. 
	
\end{abstract}


\section{Introduction}

\IEEEPARstart{R}{ecently}, intelligent reconfigurable surface (IRS) is widely studied due to the ability of controlling electromagnetic wave~\cite{ref2,ref3}, which brings a new perspective on communication and radar systems~\cite{ref4,ref5,ref6}. IRS consists of many passive reconfigurable elements, each element can achieve the desired amplitude and phase shift independently to the incident siganl by using an intelligent controller~\cite{ref7,ref8,ref9,ref10}. Therefore, IRS is endowed with the capability of arbitrarily manipulating the propagation and scattering of electromagnetic (EM) waves~\cite{ref11}. Several interesting results are presented including the generalized Snell's law, Huygens meta surface and cascaded transmit-array, which are the important aspects of array signal processing~\cite{ref12,ref13}. Taking the advantages of the IRS, IRS-based system has been proposed for the direction finding applications.

For the estimation of source direction of arrival (DOA), numerous methods have been studied for decades, such as multiple signal classification (MUSIC)~\cite{ref14} and estimating signal parameters via rotational invariance techniques (ESPRIT)~\cite{ref15}, which belong to the subspace algorithms. To achieve high estimation performance, the compressed sensing (CS) methods~\cite{ref16,ref17,ref18} have been proposed by exploiting the signal sparsity in the spatial domain, such as orthogonal matching pursuit (OMP)~\cite{ref20} and $\ell_1$-norm based singular value decomposition ($\ell_1$-SVD) ~\cite{ref21}. However, these methods discretize the spatial domain into grids, and introduce off-grid error. Hence, off-grid methods are proposed to overcome this drawback. For example, an iterative reweighted method is given in~\cite{ref22} and a sparse Bayesian inference is given in~\cite{ref23} with considering the grid-mismatch problem. Additionally, atomic norm minimization (ANM)-based methods are proposed to estimate the DOA in the continuous domain~\cite{ref24,ref25,ref26,ref27}.


However, as a specific form of norm minimization-based methods, the ANM problem is non-deterministic polynomial (NP) hard and cannot be solved efficiently~\cite{ref57}. It is usually casted to a semi-definite programming (SDP) problem after convex relaxation~\cite{ref32,ref33}. Traditional method for solving SDP problem, such as interior point method, has high computational complexity and memory requirement~\cite{ref30}. This hampers the application of ANM in practical systems.


With the development of machine learning~\cite{ref56}, the algorithms proposed in the field of machine learning can be adopted to solve the non-convex optimization efficiently. In~\cite{ref34}, two-stage and saddle-point escaping algorithms are given to escape undesired saddle points and find a local minimum efficiently, where the complexity of finding a saddle point is given in~\cite{ref29}. Additionally, a perturbing gradient descent (PGD) method is given and adds intermittent perturbations during the iterations, where the second stagnation point can be achieved without extra time~\cite{ref35}. Moreover, an iterative hard threshold (IHT) method is proposed in~\cite{ref36,ref37}, and the improved algorithms are also developed, such as normalized iterative hard thresholding (NIHT)~\cite{ref38}, conjugate gradient iterative hard thresholding (CGIHT)~\cite{ref39} and hard thresholding pursuit (HTP)~\cite{ref40}. 

Consider the ability of controlling electromagnetic waves, the IRS is applied to the direction finding system instead of the traditional radar, and the DOA estimation methods are proposed in the IRS-aided system. For example, the position perturbation in IRS-aided unmanned aerial vehicle (UAV) swarm system is considered , and a novel atomic norm-based method is proposed in~\cite{ref54}. In~\cite{ref55}, a non-iterative two-stage method is proposed for channel estimation in a multiple input and multiple-output (MIMO) system with IRS.

In this paper, a low-cost passive direction finding (LPDF) system using IRS is developted, and a non-convex based atomic norm minimization (NC-ANM) method for LPDF is proposed. The contributions of this paper are summarized below:
\begin{itemize}
	\item \textbf{A novel LPDF system is established using IRS and the corresponding signal model is formulated:} Considering the low hardware complexity and cost, a signal model reflected by IRS and received by one antenna is given, and an optimization problem is formulated for the DOA estimation;
	\item \textbf{A non-convex optimization-based method is proposed to solve the ANM problem:} A nonconvex-based ANM (NC-ANM) method is formulated via gradient threshold iteration to solve the non-convex optimization problem directly, where a threshold is adopted to keep the sparsity and a random perturbation is introduced into the iterations to escape from saddle points;
	\item \textbf{The convergence of the proposed method is analyzed:} To guarantee the performance of the proposed NC-ANM method, both the algorithm convergence and complexity are analyzed, and the Cram\'{e}r-Rao lower bound (CRLB) is derived to evaluate the estimation accuracy.
\end{itemize}

The remainder of this paper is organized as follows: The LPDF system is proposed and the received signal model is formulated in Section~\uppercase\expandafter{\romannumeral2}. The DOA estimation method based on the NC-ANM is proposed and the convergence analysis is carried out in Section~\uppercase\expandafter{\romannumeral3}. The CRLB of DOA estimation is given in Section~\uppercase\expandafter{\romannumeral4}, and the simulation results are shown in Section~\uppercase\expandafter{\romannumeral5}. Finally, Section~\uppercase\expandafter{\romannumeral6} concludes the paper.

\emph{Notation:} Lowercase and uppercase bold letters represent vector and matrix, such as $\boldsymbol{a}$ and $\boldsymbol{A}$. $(\cdot)^{\text{T}}$ and $(\cdot)^{\text{H}}$ denote the matrix transpose and the Hermitian transpose, respectively. $(\cdot)^{*}$ denotes the conjugate. $\left\|{\cdot}\right\|_0$, $\left\|{\cdot}\right\|_1$ and $\left\|{\cdot}\right\|_2$ denote the $\ell_0$ norm, the $\ell_1$ norm and the $\ell_2$ norm, respectively. $\left\|{\cdot}\right\|_{\text{F}}$ denotes the Frobenius norm. $\odot$ denotes the Hadamard production. $\text{Tr}\{\cdot\}$ denotes the trace of a matrix. $\text{Re}\{\cdot\}$ denotes the real part of a complex number. $\text{Var}\{\cdot\}$ denotes the variance and $\text{vec}\{\cdot\}$ denotes the vectorization.

\section{The Low-Cost Passive Direction Finding SYSTEM}
\begin{figure}[!t]
	\centering
	\includegraphics[width=1\columnwidth]{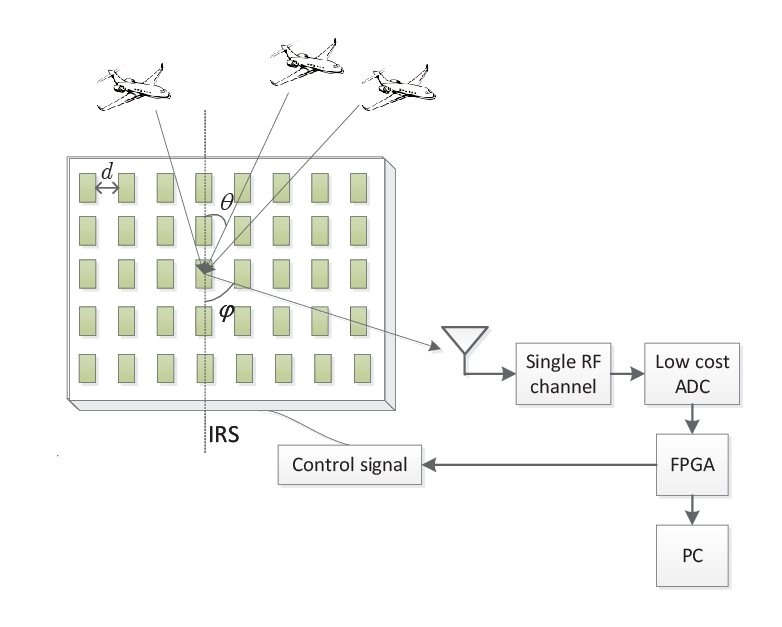}
	\caption{The LPDF system block.}
	\label{fig.1}
\end{figure}

Considering the low-cost and low-complexity, a LPDF system is formulated by an IRS, a receiver antenna, and a field programmable gate array (FPGA), as shown in Fig.~\ref{fig.1}. Different from the traditional direction finding systems, where multiple receiving channels are used to measure the different delays for the direction estimation, the proposed LPDF system only needs one full-functional receiving channel. The incident signals on the IRS are reflected to the receiver. To achieve this operation, the code sequences such as "01001100…" are set in advance, and the corresponding voltages generated by the FPGA are provided to the IRS. Thus, each “0” or “1” element of IRS can be realized by controlling the reflected phase as 0 or $\pi$ with FPGA. With the different code sequences, the reflected signals with the different phase shifts are received by the antenna. 

Assuming that the IRS consists of $N$ elements, and the position of the $n$-th $ (n=0,1,...,N-1) $ element is denoted as $ {d}_{n} $. The IRS generates specific amplitude and phase shift using the FPGA controller to reflect the signals to a specified direction. At this direction $\varphi$, a receiver is placed to receive the reflected signals. Considering the DOA estimation problem for $K$ far-field signals, the $k$-th signal from the direction $\theta_{k}$ is denoted as ${{s}_{k}}(t)$ $(k=0,1,...,K-1)$, where $t$ is the time. The bandwidth of the signals is narrow, while the sampling frequency of IRS is high, which remains signals unchanged during the multiple measurements.

First, the signals received by the $n$-th element of IRS can be expressed as 
\begin{equation}
\label{deqn_ex1a}
{{r}_{n}}(t)=\sum\limits_{k=0}^{K-1}{{{s}_{k}}(t){{e}^{j2\pi \frac{{{d}_{n}}}{\lambda }\sin {{\theta }_{k}}}}},
\end{equation}
where $\lambda$ denotes the wavelength. During the $p$-th $(p=0,1,...,P-1)$ measurement, the reflected signal can be formulated as
\begin{align}
	\begin{split}
		x_p(t)= & {\sum\limits_{n=0}^{N-1}{A_{n,p} e^{j\phi_{n,p}}}}{r_n(t)}{{e}^{j2\pi \frac{{{d}_{n}}}{\lambda }\sin \varphi }}.
	\end{split}
\end{align}	
where $A_{n,p}$ and $\phi_{n,p}$ denote the amplitude and phase shift of the $n$-th element, respectively. Collect signals of $P$ measurements into a matrix
\begin{align}
\begin{split}
  		& \boldsymbol{X}\triangleq {{[\boldsymbol{x}_1,\boldsymbol{x}_2,...,\boldsymbol{x}_{P-1}]}^\text{T}} ,
\end{split}
\end{align}
where ${\boldsymbol{x}_p} = [x_p(0),x_p(1),...,x_p(t-1)]$. The steering matrix is defined as
\begin{align}
	\begin{split}
		\boldsymbol{A}\triangleq [\boldsymbol{a}({{\theta }_{0}}),\boldsymbol{a}({{\theta }_{1}}),...,\boldsymbol{a}({{\theta }_{K-1}})],
	\end{split}
\end{align}
where the $k$-th column $\boldsymbol{a}({{\theta }_{k}})$ denotes a steering vector
\begin{align}
	\begin{split}
		\boldsymbol{a}({{\theta }_{k}})\triangleq {{\left[ \begin{matrix}
					{{e}^{j2\pi \frac{{{d}_{0}}}{\lambda }\sin {{\theta }_{k}}}},...,{{e}^{j2\pi \frac{{{d}_{N-1}}}{\lambda }\sin {{\theta }_{k}}}} \\ 
				\end{matrix} \right]}^{\text{T}}}.
	\end{split}
\end{align}
The signal matrix is defined as
\begin{align}
	\begin{split}
		{\boldsymbol{S}\triangleq {{[{\boldsymbol{s}_{0}},{\boldsymbol{s}_{1}},...,{\boldsymbol{s}_{K-1}}]}^{\text{T}}}},
	\end{split}
\end{align}
where the $k$-th signal ${\mathbf{s}_{k}}={{\left[{{s}_{k}}(0),{{s}_{k}}(1)..., {{s}_{k}}(t-1) \right]}^{\text{T}}}$. Based on the definitions, the received signal with additive noise can be written as
\begin{align}
	\begin{split}
		\boldsymbol{Y} = \boldsymbol{X} + \boldsymbol{W} =  {{\boldsymbol{B}}^{\text{T}}}\boldsymbol{Z+W},
	\end{split} \label{eq8}
\end{align}
where 
${\boldsymbol{Y}\triangleq {\left [{\boldsymbol{y}_{0}},{\boldsymbol{y}_{1}},...,{\boldsymbol{y}_{P-1}}\right ]}}$, 
$\boldsymbol{Z}=\boldsymbol{AS}=\left [{\boldsymbol{z}_{0}},{\boldsymbol{z}_{1}},...,{\boldsymbol{z}_{P-1}}\right ]$. $\boldsymbol{y}_{p}$ and $\boldsymbol{z}_{p}$ denote the $p$-th column vector of $\boldsymbol{Y}$ and $\boldsymbol{Z}$.
$\boldsymbol{w}=\text{vec}\left\{ \boldsymbol{W} \right\}\sim \mathcal{C}\mathcal{N}(0,\sigma _{n}^{2}\boldsymbol{I})$, and $\sigma _{n}$ is the variance of noise. The measurement matrix $\boldsymbol{B}$ is defined as
\begin{align}
	\begin{split}
		{\boldsymbol{B}\triangleq [\boldsymbol{b}_0,\boldsymbol{b}_1,...,\boldsymbol{b}_{P-1}]},
	\end{split}
\end{align}
where we have
\begin{align}
	\begin{split}
		\boldsymbol{a}(\varphi )\triangleq {{\left[ \begin{matrix}
					{{e}^{j2\pi \frac{{{d}_{0}}}{\lambda }\sin \varphi }},...,{{e}^{j2\pi \frac{{{d}_{N-1}}}{\lambda }\sin \varphi }} 
				\end{matrix} \right]}^{\text{T}}},
	\end{split}
\end{align}
\begin{align}
	\begin{split}
		\boldsymbol{e}(p) \!\triangleq\!{{\left[ \begin{matrix}
					{{A}_{0,p}}{{e}^{j{{\phi }_{0,p}}}},...,{{A}_{N-1,p}}{{e}^{j{{\phi }_{N-1,p}}}}
				\end{matrix} \right]}^{\text{T}}}.
	\end{split}
\end{align}
\begin{align}
	\begin{split}
		\boldsymbol{b}_p & \triangleq \boldsymbol{a}(\varphi )\odot \boldsymbol{e}(p) \\ & =\left [	{{A}_{0,p}}{{e}^{j(2\pi \frac{{{d}_{0}}}{\lambda }\sin \varphi+{{\phi }_{0,p})}}}, \right.\\& \left.   ~~~~~~...,{{A}_{N-1,p}}{{e}^{j(2\pi \frac{{{d}_{N-1}}}{\lambda }\sin \varphi+{{\phi }_{N-1,p})}}}\right ] .
	\end{split}
\end{align}

After getting the received signal $\boldsymbol{Y}$ and the measurement matrix $\boldsymbol{B}$, we aim to estimate the DOA parameters $\boldsymbol{\theta}\triangleq \left[\theta_0,\theta_1,\dots,\theta_{K-1}\right]^{\text{T}}$ by exploiting the signal sparsity in the continuous domain. To avoid the approximation of convex relaxation and high computational complexity, we propose a novel nonconvex method to solve the original optimization problem of atomic norm.

\section{The NC-ANM Based Method for DOA Estimation  With LPDF System}
\subsection{Traditional ANM Method for DOA Estimation}

In the traditional ANM-based method, the DOA estimation in (\ref{eq8}) is transferred into
\begin{align}
	\begin{split}
		\underset{\boldsymbol{Z}}{\mathop{\min }}\,\frac{1}{2}\left\| \boldsymbol{Y-}{{\boldsymbol{B}}^{\text{T}}}\boldsymbol{Z} \right\|_2^2+\tau {{\left\| \boldsymbol{Z} \right\|}_{\mathcal{A}}},\label{eq13}
	\end{split}
\end{align}
where $\tau$ denotes the regularization parameter and is uesd to control the ballance between the reconstruction performance and the sparsity \cite{ref27}. ${{\left\| \boldsymbol{Z} \right\|}_{\mathcal{A}}}$ denotes the atomic norm of $\boldsymbol{Z}$, and is defined as
\begin{align}
	\begin{split}
		{{\left\| \boldsymbol{Z} \right\|}_{\mathcal{A}}}\triangleq \inf \Bigg\{ &\left. \sum\limits_{k}{{{c}_{k}}}:\boldsymbol{Z}=\sum\limits_{k}{{{c}_{k}}}\boldsymbol{a}({{\theta }_{k}}){\boldsymbol{d}_{k}^{\text{T}}}, \right.\\
		&\left.{{{c}_{k}}\ge 0,{{\left\| {{\boldsymbol{d}}_{k}} \right\|}_{2}}=1}\right.\Bigg\}. \label{eq14}
	\end{split}
\end{align}
Then, (\ref{eq14}) can be approximately casted to a  SDP problem \cite{ref41}  
\begin{align}
	\begin{split}
		\underset{\boldsymbol{Z,u,V}}{\mathop{\min }}\,
		&~~~\text{Tr}(\boldsymbol{T}(\boldsymbol{u}))+\text{Tr}(\boldsymbol{V})\\ 
		s.t.&~~~\left[ \begin{matrix}
			\boldsymbol{T}(\boldsymbol{u}) & {\boldsymbol{Z}}  \\
			\boldsymbol{Z}^{\text{H}} & \boldsymbol{V} \\
		\end{matrix} \right]\succeq 0 , \label{eq15}\\
	\end{split}
\end{align}
where ${\boldsymbol{T(u)}=\sum\limits_{k}{{{c}_{k}}\boldsymbol{a}({{\theta }_{k}}){{\boldsymbol{a}}^{\text{H}}}({{\theta }_{k}})}}\in {\mathbb{C}^{N\times N}}$, $\boldsymbol{u}=\sum\limits_{k}{{{c}_{k}}\boldsymbol{a}({{\theta }_{k}})}$ is the first column of ${\boldsymbol{T(u)}}$, and ${\boldsymbol{V}=\sum\limits_{k}{{c}_{k}{\boldsymbol{d}_{k}}{\boldsymbol{d}_{k}}^{\text{H}}}}$. The optimization problem (\ref{eq15}) can be solved by the CVX toolbox in MATLAB directly~\cite{ref42}.

However, this method has two issues. (1) ${{\left\| \boldsymbol{Z} \right\|}_{\mathcal{A}}}$ is used as an approximation of ${{\left\| \boldsymbol{Z} \right\|}_{\mathcal{A},0}}$ in (\ref{eq13}) since the ${{\ell}_{\mathcal{A},0}}$ norm minimization problem is NP hard and unfeasible to be computed~\cite{ref43,ref44}, where ${{\left\| \boldsymbol{Z} \right\|}_{\mathcal{A},0}}$ is defined as
\begin{align}
	\begin{split}
		{{\left\| \boldsymbol{Z} \right\|}_{\mathcal{A},0}}\triangleq \underset{K}{\mathop{\inf }}\, \Bigg\{ &\left. \boldsymbol{Z}=\sum\limits_{k=0}^{K-1}{{{c}_{k}}}\boldsymbol{a}({{\theta }_{k}}){\boldsymbol{d}_{k}^{\text{T}}}, \right.\\
		&\left.{{{c}_{k}}\ge 0,{{\left\| {{\boldsymbol{d}}_{k}} \right\|}_{2}}=1}\right.\Bigg\}.
	\end{split}
\end{align}
This will lead to inaccuracy. (2) Solving the SDP problem in (\ref{eq15}) by CVX toolbox has high computational complexity. Therefore, we proposed a direct method to solve the non-convex ANM optimization problem with low complexity.

\subsection{DOA Estimation Based on NC-ANM Method}

In the DOA estimation problem (\ref{eq13}),  ${{\left\| \boldsymbol{Z} \right\|}_{\mathcal{A},0}}$ is adopted directly  instead of the convex relaxation
\begin{align}
	\begin{split}
		\underset{\boldsymbol{Z}}{\mathop{\min }}\,\frac{1}{2}\left\| \boldsymbol{Y}-{{\boldsymbol{B}}^{\text{T}}}\boldsymbol{Z} \right\|_{{2}}^{2}+\tau {{\left\| \boldsymbol{Z} \right\|}_{\mathcal{A},0}}.
	\end{split}\label{eq17}
\end{align}
To solve the optimization problem in (\ref{eq17}), a gridless DOA estimation method ANM based on non-convex optimization is proposed instead of convex relaxation. The proposed NC-ANM method includes two steps, i.e., the iteration step and the threshold step. In the iteration step, the iterative regulation is formulated, where the gradients are drived and a random perturbation is added to avoid converging to saddle points. In the threshold step, the sparsity of $c_k$ is limited by comparing with the threshold, which can improve the convergence speed and guarantee the conditional restriction of ${{\left\| \boldsymbol{Z} \right\|}_{\mathcal{A},0}}$. 

First, the optimization problem can be rewritten as
\begin{align}
	\begin{split}
\underset{c_k,\boldsymbol{\beta}_k,{\theta}_k}{\mathop{\min }}\,\frac{1}{2}F(c_k,\boldsymbol{\beta}_k,{\theta}_k)+\tau {{\left\| \boldsymbol{Z} \right\|}_{\mathcal{A},0}}, \label{eq_non}
	\end{split}
\end{align}

 \begin{align}
	\begin{split}
		F(c_k,\boldsymbol{\beta}_k,{\theta}_k)=& \left\| \boldsymbol{Y}-{\boldsymbol{B}}^{\text{T}}{\sum\limits_{k}{{c}_{k}}}{\boldsymbol{a}({\theta }_{k}){e^{j \boldsymbol{\beta}_k}}} \right\|_{{2}}^{2},  \label{eq18}
	\end{split}
\end{align}
where (\ref{eq_non}) is a nonconvex problem. ${e^{j \boldsymbol{\beta}_k}}$ can be treated as $\boldsymbol{d}_k^\text{T}$ in definition of the atomic set. The coefficients $c_k$ ($k=0,1,\dots,K-1$) are sparse and ${{\theta }_{k}}$ is an unknown angle to be estimated in the continuous domain. In order to solve the nonconvex problem effectively using the iterative method, the gradients of $c_k$, $\boldsymbol{\beta}_k$, ${{\theta }_{k}}$ are calculated as
\begin{align}
	\begin{split}
		&\nabla_{c_k} F = \frac{\partial(F)}{\partial {c_k}} \\ 
		=& 2\text{Re} \left\{\left[\boldsymbol{Y}-{\boldsymbol{B}}^{\text{T}}{\sum\limits_{k}{{c}_{k}}}{\boldsymbol{a}({{\theta }_{k}})}{e^{j\boldsymbol{\beta}_k}}\right]^{\text{H}}\left[-{\boldsymbol{B}}^{\text{T}}{\boldsymbol{a}({{\theta }_{k}})}{e^{j\boldsymbol{\beta}_k}}\right]\right\}.  
	\end{split}
\end{align}
Similarly, we can get
\begin{align}
	\begin{split}
		\nabla_{\boldsymbol{\beta}_k} F =& \frac{\partial(F)}{\partial {\boldsymbol{\beta}_k}} \\
		= & 2\text{Re} \left\{\left[\boldsymbol{Y}-{\boldsymbol{B}}^{\text{T}}{\sum\limits_{k}{{c}_{k}}}{\boldsymbol{a}({{\theta }_{k}})}{e^{j\boldsymbol{\beta}_k}}\right]^{\text{H}} \right.\\ & ~~~~~~~~~~~~\left. \left[-j{\boldsymbol{B}}^{\text{T}}{\sum\limits_{k}{{c}_{k}}}{\boldsymbol{a}({\theta }_{k})}{e^{j\boldsymbol{\beta}_k}}\right]\right\},   
	\end{split}
\end{align}
\begin{align}
	\begin{split}
		\nabla_{{\theta }_{k}} F = &\frac{\partial(F)}{\partial {{\theta }_{k}}} \\
		= & 2\text{Re} \left\{\left[\boldsymbol{Y}-{\boldsymbol{B}}^{\text{T}}{\sum\limits_{k}{{c}_{k}}}{\boldsymbol{a}({{\theta }_{k}})}{e^{j\boldsymbol{\beta}_k}}\right]^{\text{H}} \right.\\ & ~~~~~~~~~~~~\left. \left[-{\boldsymbol{B}}^{\text{T}}{\sum\limits_{k}{{c}_{k}}}{\boldsymbol{\gamma}}{\boldsymbol{a}({{\theta }_{k}})}{e^{j\boldsymbol{\beta}_k}}\right]\right\},
	\end{split}
\end{align}
where ${\boldsymbol{\gamma}}=\left[0, j\pi,...,j(N-1)\pi\right]^{\text{T}}$ with the space of IRS elements being $\frac{1}{2}\lambda$.

With setting the sparsity, the sparse vectors $\boldsymbol{c}$, $\boldsymbol {\beta}$, $\boldsymbol{\theta}$ and corresponding gradients can be obtained, which determines the iterative regulation. Here, ${{c}_{k}}$, $\boldsymbol{\beta}_k$, and $\theta_{k}$ represent the non-zero elements in $\boldsymbol{c}$, $\boldsymbol {\beta}$, and $\boldsymbol{\theta}$. It is noteworthy that the selection of sparsity has to compromise the estimation accuracy and computational complexity.

The details about the proposed algorithm are shown in Algorithm~\ref{nc-anm}.
 
\begin{algorithm}
	\caption{NC-ANM Method} \label{nc-anm}
	\begin{algorithmic}[1]
		\STATE  \emph{Input:} The received signal $\boldsymbol Y$, the measurement matrix $\boldsymbol{B}$, the maximum number of iterations $Q$, the step size $\eta$, the threshold $T$ and the sparsity $S$.
		\STATE \emph{Initialization:} The random initialization $\boldsymbol{c}^{0} \in \mathbb{C}^{S}$, $\boldsymbol {\beta}^{0}\in \mathbb{C}^{S}$, $\boldsymbol{\theta}^{0}\in \mathbb{C}^{S}$ and $q=0$.
		\WHILE{$q<Q$}
		\STATE Take the iteration as \begin{align}
			{\boldsymbol{c}^{q+1}}={\boldsymbol{c}^{q}}-\eta \nabla_{\boldsymbol{c}^{q}} F,
		\end{align} 
	    \begin{align}
	  	    {\boldsymbol{\beta}^{q+1}}={\boldsymbol {\beta}^{q}}-\eta \nabla_{\boldsymbol {\beta}^{q}} F,
	    \end{align}
	    \begin{align}
	    	{\boldsymbol{\theta}^{q+1}}={\boldsymbol{\theta}^{q}}-\eta \nabla_{\boldsymbol{\theta}^{q}} F,
	    \end{align} 
    where $\boldsymbol{c}^{q}$, $\boldsymbol{\beta}^{q}$, and $\boldsymbol{\theta}^{q}$ denote the updated $\boldsymbol{c}$, $\boldsymbol{\beta}$, and $\boldsymbol{\theta}$ in the $q$-th iteration.
	    \IF {perturbation condition $\nabla F(\cdot) \le \epsilon$ holds}
	    \STATE \begin{align}
	    	{\boldsymbol{c}^{q+1}}={\boldsymbol{c}^{q+1}}+\boldsymbol{\xi^q},\label{eq31}
	    \end{align}
        \begin{align}
        	{\boldsymbol{\beta}^{q+1}}={\boldsymbol{\beta}^{q+1}}+\boldsymbol{\xi^q}, \label{eq32}
        \end{align}
        \begin{align}
        	{\boldsymbol{\theta}^{q+1}}={\boldsymbol{\theta}^{q+1}}+\boldsymbol{\xi^q}, \label{eq33}
        \end{align}
        \ENDIF	    
		\IF {${c}^{q+1}_{s}\geq T$}
	    \STATE {\begin{align}
			{{c}^{q+1}_{s}={c}^{q+1}_{s}},
		\end{align}}where ${c}^{q+1}_{s}$ denotes the $s$-th entry of $\boldsymbol{c}^{q+1}$. This step keeps all elements larger than the threshold.
	    \ELSE 
	    \STATE Set the values of remaining elements be $0$  {\begin{align}
	    	{{c}^{q+1}_{s}}=0,
	    \end{align}} which ensure the sparsity of $\boldsymbol{c}$ and ${{\left\| \boldsymbol{Z}^q \right\|}_{\mathcal{A},0}}\leq {{\left\| \boldsymbol{Z}^{q+1} \right\|}_{\mathcal{A},0}}$.
        \ENDIF
        \STATE Arrange $\boldsymbol{{c}}^{q+1}$, $\boldsymbol{\beta}^{q+1}$, $\boldsymbol{{\theta}}^{q+1}$ in descending order.
        \IF{the element of $\boldsymbol{{\theta}}^{q+1}$ satisifies
        \begin{align}
        	\left| {{\theta}^{q+1}_{s+1}-{\theta}^{q+1}_s} \right| \le \frac{1}{S} or \left| {{\theta}^{q+1}_{s}} \right| >90^{\circ}
        \end{align}}
        \STATE set ${{c}^{q+1}_{s}}=0$. Thus $\boldsymbol{\tilde{c}}^{q+1}$ is obtained after nonlinear transformation.
        \ENDIF
		\STATE $q\leftarrow q+1$.
		\ENDWHILE 
		\STATE \emph{Output:} The DOA estimation $\boldsymbol{\hat{\theta}}$ can be obtained. 
	\end{algorithmic}
\end{algorithm}
In ({\ref{eq31}})-({\ref{eq33}}), we add a random perturbation when the gradient is suitably small, which aims to avoid converging to the saddle point in non-convex optimization. The elements of perturbation matrix $\boldsymbol{\xi}$ are sampled uniformly from a sphere with a zero center and radius  $\tilde{\xi }$. Taking $\boldsymbol{c}$ as an example, we can rewrite the update rule as
\begin{equation}
	\begin{split}
 {\boldsymbol{c}^{q+1}} & ={{\boldsymbol{c}}^{q}}-\eta \nabla_{\boldsymbol{c}^{q}} F \\ & ={{\boldsymbol{c}}^{q}}-\eta \nabla_{\boldsymbol{c}^{q}} \tilde{F}-\eta \underbrace{(\nabla_{\boldsymbol{c}^{q}} F-\nabla_{{\boldsymbol{c}^{q}}} \tilde{F})}_{\boldsymbol{R}({{\boldsymbol{c}}^{q}})},	
 \end{split}
\end{equation}
where $\nabla_{\boldsymbol{c}} \tilde{F}$ satisfies $\nabla_{\boldsymbol{c}} \tilde{F}=\mathbb{E}[\nabla_{\boldsymbol{c}} F]$. Assuming the iteration ${{\boldsymbol{c}}^{q}}$ is independent that we can control the size of perturbation term according to the central limit theorem (CLT) \cite{ref45}
\begin{align}
	\begin{split}
		\left| \boldsymbol{R}({{\boldsymbol{c}}^{q}}) \right|= & \left|\nabla_{\boldsymbol{c}^{q}} F-\nabla_{{\boldsymbol{c}^{q}}} \tilde{F}\right| \\
		\lesssim & \sqrt{\text{Var}(\boldsymbol{R}({{\boldsymbol{c}}^{q}}))\text{polylog}(P)} \\ \lesssim& \sqrt{\frac{\text{polylog} (P)}{P}}.
	\end{split}
\end{align}
where $\text{polylog}\{\cdot\}$ denotes the multiple logarithm in statistics~\cite{ref53, ref58}. Then we can get the range of $\tilde{\xi }$ as
\begin{align}
	\begin{split}
		\tilde{\xi }\lesssim \eta \sqrt{\text{Var}(\boldsymbol{R}({{\boldsymbol{c}}^{q}}))\text{polylog}(P)}\lesssim \eta \sqrt{\frac{\text{polylog} (P)}{P}}.
	\end{split}
\end{align}

\subsection{The Analysis of Algorithm Convergence}

To analyse the convergence of the proposed method, we give the following Definition~{\ref{Def1}} and Proposition~{\ref{Proposition1}}, which have been proved in Appendix~{\ref{appA}}. 

\begin{myDef}\label{Def1}
For the gradient Lipschitz continuous function $f$, only when $0\leq l \leq L$, the following inequality holds for any $x,y \in \mathbb{R}^n$: 
\begin{align}
	\begin{split}
	&\nabla f(y)^{\text{T}}(x-y)+\frac{l}{2}\left\|{x-y}\right\|^2 \\ & \leq f(x)-f(y) \\ & \leq \nabla f(y)^{\text{T}}(x-y)+\frac{L}{2}\left\|{x-y}\right\|^2. \label{eq35}
	\end{split} 
\end{align} 
where $l$ denotes the Lipschitz constant and $L$ is the upper bound of $\nabla f$.
\end{myDef} 

Here, (\ref{eq18}) can easily satisfy the condition of gradient Lipschitz continuous by designing measurement matrix ${\boldsymbol{B}}$. It aims to limit the gradient.
\begin{myProposition}{\label{Proposition1}}
When function $f$ satisfies the Definition~{\ref{Def1}}, let $0 \leq \zeta < \frac{2l}{L}$, $\rho= \sqrt{1-2\zeta l + \zeta^2L^2}$ we can get
\begin{align}
	\begin{split}
		f(x)-f(y) \leq \langle{f(y),x-y} \rangle +\frac{1+\rho}{2\zeta}\left\|{x-y}\right\|^2. \label{eq36}
	\end{split} 
\end{align}
\end{myProposition}

Assume that the measurement matrix $\boldsymbol{B}$ is known with designing the codes of IRS, and the gradient in algorithm $\nabla F$ satisfies the Lipschitz condition. In the threshold step of $q$-th iteration, $\boldsymbol{\tilde{c}}^{q}$ is obtained after comparing $\boldsymbol{c}^{q}$ with the threshold $T$. To prove that the threshold step accelerates the iterative convergence, (\ref{eq37}) is calculated
\begin{align}
	\begin{split}
		& \left\|\boldsymbol{\tilde{c}}^{q}-\boldsymbol{c}^{q} \right\|^2_2 \\
		= & \left\|\boldsymbol{\tilde{c}}^{q}-\boldsymbol{c}^{q-1}+\eta \nabla_{\boldsymbol{c}^{q-1}} F \right\|^2_2 \\ = & \left\|\boldsymbol{\tilde{c}}^{q}-\boldsymbol{c}^{q-1} \right\|^2_2 + 2\eta \langle \nabla_{\boldsymbol{c}^{q-1}} F,\boldsymbol{\tilde{c}}^{q}-\boldsymbol{c}^{q-1} \rangle \\ & + \left\|\eta \nabla_{\boldsymbol{c}^{q-1}} F \right\|^2_2 \\ \leq & \left\|\boldsymbol{c}^{q-1}-\boldsymbol{c}^{q-1} + \eta \nabla_{\boldsymbol{c}^{q-1}} F\right\|^2_2 \\ = & \left\|\eta \nabla_{\boldsymbol{c}^{q-1}} F \right\|^2_2,
	\end{split} \label{eq37}
\end{align}
hence, we have
\begin{align}
	\begin{split}
		\langle \nabla_{\boldsymbol{c}^{q-1}} F,\boldsymbol{\tilde{c}}^{q}-\boldsymbol{c}^{q-1} \rangle \leq -\frac{1}{2\eta}\left\|\boldsymbol{\tilde{c}}^{q}-\boldsymbol{c}^{q} \right\|^2_2.
	\end{split} \label{eq38}
\end{align}
Taking $\boldsymbol{\tilde {c}}^{q}$ and $\boldsymbol{c}^{q-1}$ into the inequality in Proposition~{\ref{Proposition1}}, and combining with (\ref{eq38}), we can obtain the change of function value after once iteration step and threshold step
\begin{align}
	\begin{split}
		& F(\boldsymbol{\tilde {c}}^{q})-F(\boldsymbol{c}^{q-1}) \\ \leq & \langle \nabla_{\boldsymbol{c}^{q-1}} F,\boldsymbol{\tilde{c}}^{q}-\boldsymbol{c}^{q-1} \rangle +\frac{1+\rho}{2\zeta}\left\|\boldsymbol{\tilde{c}}^{q}-\boldsymbol{c}^{q-1} \right\|^2_2 \\ \leq & \frac{-\zeta+\eta (1+\rho)}{2\zeta \eta}\left\|\boldsymbol{\tilde{c}}^{q}-\boldsymbol{c}^{q-1} \right\|^2_2.
	\end{split}
\end{align}
By choosing the suitable step size $\eta < \frac{\zeta}{1+\rho}$, we can get that it is non-incremental from $F(\boldsymbol{c}^{q-1})$ to $F(\boldsymbol{\tilde{c}}^{q})$ and the convergence is accelerated by threshold step
\begin{align}
	\begin{split}
		F(\boldsymbol{\tilde{c}}^{q})-F(\boldsymbol{c}^{q-1}) 	\leq F(\boldsymbol{c}^{q})-F(\boldsymbol{c}^{q-1}) 
		\leq 0. \label{eq40}
	\end{split}
\end{align} 

\subsection{The Analysis of Algorithm Complexity}
For the optimization problem (\ref{eq18}), the main complexity of proposed algorithm than traditional GD lies in finding the strict saddle points and converging to the second-order stationary point, which can be formulated as
\begin{align}
	\begin{split}
		\left| \nabla F \right|\le \epsilon,\text{ }{{\lambda }_{\min }}({{\nabla }^{2}}F)\ge -\sqrt{l \epsilon},
	\end{split}
\end{align}
where a sufficient constant $\epsilon$ is adopted intead of $\nabla F=0$. We take $\boldsymbol{\theta}$ as an example. Under this condition, finding a saddle point only needs $2l (F({{\boldsymbol{\theta}}^{0}})-F(\boldsymbol{\hat{\theta}}))/{{\epsilon}^{2}}$ iterations with $\eta =1/l$, where $\boldsymbol{\hat{\theta}}$ denotes the optimal solution of $\boldsymbol{{\theta}}$. That is to say the strict saddle points can be found in $\mathcal{O}(1/{{\epsilon}^{2}})$ steeps.

The gradient satisfies the Lipschitz condition regarding smoothness, and the following formula holds
\begin{align}
	\begin{split}
		F(\boldsymbol{\theta}^{q})\leq F(\boldsymbol{\theta}^{q-1})\!+\! \nabla_{\boldsymbol{\theta}^{q-1}} F^{\text{T}}(\boldsymbol{\theta}^{q}-\boldsymbol{\theta}^{q-1})\!+\!\frac{l }{2}{{\left\| \boldsymbol{\theta}^{q}-\boldsymbol{\theta}^{q-1} \right\|}^{2}}\label{eq42}.
	\end{split}
\end{align}
Taking $\boldsymbol{\theta}^{q}=\boldsymbol{\theta}^{q-1}-\eta \nabla_{\boldsymbol{\theta}^{q-1}} F$ into (\ref{eq42}) and limiting~$0<\eta \le \frac{1}{l }$, we can get
\begin{align}
	\begin{split}
		F(\boldsymbol{\theta}^{q})\le & F(\boldsymbol{\theta}^{q-1})-(1-\frac{\eta l }{2})\eta {{\left\| \nabla_{\boldsymbol{\theta}^{q-1}} F \right\|}^{2}} \\ \le & F(\boldsymbol{\theta}^{q-1})-\frac{\eta}{2} {{\left\| \nabla_{\boldsymbol{\theta}^{q-1}} F \right\|}^{2}} \\ \le & F(\boldsymbol{\hat{\theta}})+\nabla_{\boldsymbol{\theta}^{q-1}}F^{\text{T}}(\boldsymbol{\theta}^{q-1}-\boldsymbol{\hat{\theta}})-\frac{\eta}{2} {{\left\| \nabla_{\boldsymbol{\theta}^{q-1}} F \right\|}^{2}},\label{eq43}
	\end{split}
\end{align}
then taking $\nabla_{\boldsymbol{\theta}^{q-1}} F=\frac{1}{\eta}(\boldsymbol{\theta}^{q}-\boldsymbol{\theta}^{q-1})$ into (\ref{eq43}), the following inequality can be obtained
\begin{align}
	\begin{split}
		F({\boldsymbol{\theta}^{q}})\le F(\boldsymbol{\hat{\theta}})+\frac{1}{2\eta }({{\left\| \boldsymbol{\theta}^{q-1}-\boldsymbol{\hat{\theta}}\right\|}^{2}}-{{\left\| {{\boldsymbol{\theta}}^{q}}-\boldsymbol{\hat{\theta}} \right\|}^{2}}).
	\end{split}
\end{align}
When $q=1,2,...,Q$, sum all iterations to get
\begin{align}
	\begin{split}
		 \sum\limits_{q=1}^{Q}(F({{\boldsymbol{\theta}}^{q}}))-QF(\boldsymbol{\hat{\theta}}) &\le \frac{1}{2\eta }({{\left\| {{\boldsymbol{\theta}}^{0}}-\boldsymbol{\hat{\theta}} \right\|}^{2}}-{{\left\| {{\boldsymbol{\theta}}^{Q}}-\boldsymbol{\hat{\theta}} \right\|}^{2}}) \\ 
		& \le \frac{1}{2\eta }{{\left\| {{\boldsymbol{\theta}}^{0}}-\boldsymbol{\hat{\theta}} \right\|}^{2}}.
	\end{split}
\end{align} 
 The inequality  $F(\boldsymbol{c}^{q})-F(\boldsymbol{c}^{q-1}) \leq 0$ has been given in (\ref{eq40}), thus $QF({{\boldsymbol{\theta}}^{Q}}) \leq \sum\limits_{q=1}^{Q}(F({{\boldsymbol{\theta}}^{q}}))$. Finally, we can get
\begin{align}
	\begin{split}
		 F({{\boldsymbol{\theta}}^{Q}})-F(\boldsymbol{\hat{\theta}}) 
		\le \frac{1}{2\eta Q}{{\left\| {{\boldsymbol{\theta}}^{0}}-\boldsymbol{\hat{\theta}} \right\|}^{2}}.
	\end{split}
\end{align}
That means the convergence rate is $\mathcal{O}(1/Q)$. Under Hessian-Lipschitz condition, converging to a second-order-stationary point costs almost the same time as GD converging to a first order-stationary point. Usually, we bound the numbers of iterations and find $F({{\boldsymbol{\theta}}^{Q}})-F(\boldsymbol{\hat{\theta}})\le \epsilon$, which only needs $\mathcal{O}(1/\epsilon)$ steps. That is much lower in complexity than using CVX, and more stable in performance than solving (\ref{eq13}) directly by the least square (LS) method\cite{ref48}.

\section{CRLB For DOA Estimation With IRS System}
In this section, the CRLB will be derived to show the performance of DOA estimation problem. First, we assume that $\boldsymbol{s}$ follows the zeros mean Gaussian distribution with $\boldsymbol{s}=\text{vec}\left\{ \boldsymbol{S} \right\}\sim \mathcal{C}\mathcal{N}(0,\boldsymbol{D})$, where $\mathcal{E}(\boldsymbol{s}{{\boldsymbol{s}}^{\text{H}}})=\boldsymbol{D}$. Then the received signal can be written as
\begin{align}
	\begin{split}
		\boldsymbol{y}\triangleq \text{vec}\left\{ \boldsymbol{Y} \right\}=(\boldsymbol{I}\otimes {{\boldsymbol{B}}^{\boldsymbol{T}}}\boldsymbol{A})\boldsymbol{s}+\boldsymbol{w},
	\end{split}
\end{align}
It can be gotten that $\boldsymbol{y}$ follows the Gaussian distribution $\boldsymbol{y}\sim \mathcal{C}\mathcal{N}(0,\boldsymbol{G})$, where 
\begin{align}
	\boldsymbol{G}\triangleq (\boldsymbol{I}\otimes {{\boldsymbol{B}}^{\boldsymbol{\text{T}}}}\boldsymbol{A})\boldsymbol{D}[\boldsymbol{I}\otimes {{({{\boldsymbol{B}}^{\boldsymbol{\text{T}}}}\boldsymbol{A})}^{\text{H}}}]+\sigma _{n}^{2}\boldsymbol{I}.
\end{align}

Then the probability density function can be expressed as
\begin{align}
	\begin{split}
		f(\boldsymbol{x})=\frac{1}{{{\pi }^{P}}\det \left\{ \boldsymbol{G} \right\}}{{e}^{-{{{\boldsymbol{y}}^{\text{H}}}{{\boldsymbol{G}}^{-1}}\boldsymbol{y}}}}.
	\end{split}
\end{align}
The Fisher information matrix $\boldsymbol{F}$ can be written as
\begin{align}
	\begin{split}
		\boldsymbol{F}\triangleq \left[ \begin{matrix}
			{{\boldsymbol{F}}_{1,1}} & {{\boldsymbol{F}}_{1,2}}  \\
			{{\boldsymbol{F}}_{2,1}} & {{\boldsymbol{F}}_{2,2}}  \\
		\end{matrix} \right],
	\end{split}
\end{align}
where we have
\begin{align}
	\begin{split}
		{{\boldsymbol{F}}_{1,1}}=-\mathcal{E}\left\{ \left. \frac{\partial \ln f(\boldsymbol{y};\boldsymbol{\theta },\boldsymbol{\psi })}{\partial \boldsymbol{\theta }\partial \boldsymbol{\theta }} \right|\boldsymbol{\theta },\boldsymbol{\psi } \right\},
	\end{split}
\end{align}
$\boldsymbol{\psi }$ denotes the all unknown information except $\boldsymbol{\theta }$. Then the ${{k}_{1}}$,  ${{k}_{2}}$-th entry of Fisher information matrix ${{\boldsymbol{F}}_{1,1}}$ can be obtained as
\begin{align}
	\begin{split}
		\boldsymbol{F}_{{{k}_{1}},{{k}_{2}}}^{1,1}=\frac{\partial \ln \det \left\{ \boldsymbol{G} \right\}}{\partial {{\theta }_{{{k}_{1}}}}\partial {{\theta }_{{{k}_{2}}}}}+\mathcal{E}\left\{ \frac{\partial {{{\boldsymbol{y}}}^{\text{H}}}{{\boldsymbol{G}}^{-1}}\boldsymbol{y})}{\partial {{\theta }_{{{k}_{1}}}}\partial {{\theta }_{{{k}_{2}}}}} \right\},
	\end{split}
\end{align}
and each term can be obtained as
\begin{align}
	\begin{split}
		& \frac{\partial \ln \det \left\{ \boldsymbol{G} \right\}}{\partial {{\theta }_{{{k}_{1}}}}\partial {{\theta }_{{{k}_{2}}}}}=\text{Tr}\left\{ \frac{\partial {{\boldsymbol{G}}^{-1}}\frac{\partial \boldsymbol{G}}{\partial {{\theta }_{{{k}_{1}}}}}}{\partial {{\theta }_{{{k}_{2}}}}} \right\} \\ 
		& =\text{Tr}\left\{ \frac{\partial {{\boldsymbol{G}}^{-1}}}{\partial {{\theta }_{{{k}_{2}}}}}\frac{\partial \boldsymbol{G}}{\partial {{\theta }_{{{k}_{1}}}}} \right\}+\text{Tr}\left\{ {{\boldsymbol{G}}^{-1}}\frac{\partial \boldsymbol{G}}{\partial {{\theta }_{{{k}_{1}}}}\partial {{\theta }_{{{k}_{2}}}}} \right\} ,
	\end{split}
\end{align}
\begin{align}
	\begin{split}
		\mathcal{E}\left\{ \frac{\partial {{{\boldsymbol{y}}}^{\text{H}}}{{\boldsymbol{G}}^{-1}}\boldsymbol{y})}{\partial {{\theta }_{{{k}_{1}}}}\partial {{\theta }_{{{k}_{2}}}}} \right\}=\text{Tr}\left\{ \frac{\partial {{\boldsymbol{G}}^{-1}}}{\partial {{\theta }_{{{k}_{1}}}}\partial {{\theta }_{{{k}_{2}}}}}\boldsymbol{G} \right\},
	\end{split}
\end{align}
where $\frac{\partial \boldsymbol{G}}{\partial {{\theta }_{k}}}$, $\frac{\partial \boldsymbol{G}}{\partial {{\psi }_{k}}}$ can be calculated as
\begin{align}
	\begin{split}
		& \frac{\partial \boldsymbol{G}}{\partial {{\theta }_{k}}}=(\boldsymbol{I}\otimes {{\boldsymbol{B}}^{\text{T}}}\frac{\partial \boldsymbol{A}}{\partial {{\theta }_{k}}})\boldsymbol{D}(\boldsymbol{I}\otimes {{({{\boldsymbol{B}}^{\text{T}}}\boldsymbol{A})}^{\text{H}}}) \\ 
		& ~~~~~~+(\boldsymbol{I}\otimes {{\boldsymbol{B}}^{\text{T}}}\boldsymbol{A})\boldsymbol{D}(\boldsymbol{I}\otimes {\frac{\partial \boldsymbol{A}^{\text{H}}}{\partial {{\theta }_{k}}}{{\boldsymbol{B}^{*}}})}  ,
	\end{split}
\end{align}
\begin{align}
	\begin{split}
		& \frac{\partial \boldsymbol{G}}{\partial {{\psi }_{k}}}=(\boldsymbol{I}\otimes \frac{\partial {{\boldsymbol{B}}^{\text{T}}}}{\partial {{\psi }_{k}}}\boldsymbol{A})\boldsymbol{D}(\boldsymbol{I}\otimes {{({{\boldsymbol{B}}^{\text{T}}}\boldsymbol{A})}^{\text{H}}}) \\ 
		& ~~~~~~+(\boldsymbol{I}\otimes {{\boldsymbol{B}}^{\text{T}}}\boldsymbol{A})\boldsymbol{D}(\boldsymbol{I}\otimes {\boldsymbol{A}^{\text{H}}}{\frac{\partial {{\boldsymbol{B}}^{*}}}{\partial {{\psi }_{k}}}}),
	\end{split}
\end{align}
where $\frac{\partial \boldsymbol{A}}{\partial {{\theta }_{k}}}$ and $\frac{\partial {{\boldsymbol{B}}^{\text{T}}}}{\partial {{\psi }_{k}}}$ can be obtained easily. Then we can get
\begin{align}
	\begin{split}
		\boldsymbol{F}_{{{k}_{1}},{{k}_{2}}}^{1,1}=\text{Tr}\left\{ {{\boldsymbol{G}}^{-1}}\frac{\partial \boldsymbol{G}}{\partial {{\theta }_{{{k}_{1}}}}}{{\boldsymbol{G}}^{-1}}\frac{\partial \boldsymbol{G}}{\partial {{\theta }_{{{k}_{2}}}}} \right\}.
	\end{split}
\end{align}

Similarly, we have
\begin{align}
	\begin{split}
		{{\boldsymbol{F}}_{1,2}}=-\mathcal{E}\left\{ \left. \frac{\partial \ln f(\boldsymbol{y};\boldsymbol{\theta },\boldsymbol{\psi })}{\partial \boldsymbol{\theta }\partial \boldsymbol{\psi }} \right|\boldsymbol{\theta },\boldsymbol{\psi } \right\},
	\end{split}
\end{align}
the ${{k}_{1}}$,  ${{k}_{2}}$-th entry of Fisher information matrix ${{\boldsymbol{F}}_{1,2}}$ can be obtained as
\begin{align}
	\begin{split}
		\boldsymbol{F}_{{{k}_{1}},{{k}_{2}}}^{1,2}=\frac{\partial \ln \det \left\{ \boldsymbol{G} \right\}}{\partial {{\theta }_{{{k}_{1}}}}\partial {{\psi }_{{{k}_{2}}}}}+\mathcal{E}\left\{ \frac{\partial {{{\boldsymbol{y}}}^{\text{H}}}{{\boldsymbol{G}}^{-1}}\boldsymbol{y})}{\partial {{\theta }_{{{k}_{1}}}}\partial {{\psi }_{{{k}_{2}}}}} \right\},
	\end{split}
\end{align}
and each term can be obtained as
\begin{align}
	\begin{split}
		& \frac{\partial \ln \det \left\{ \boldsymbol{G} \right\}}{\partial {{\theta }_{{{k}_{1}}}}\partial {{\psi }_{{{k}_{2}}}}}=\text{Tr}\left\{ \frac{\partial {{\boldsymbol{G}}^{-1}}\frac{\partial \boldsymbol{G}}{\partial {{\theta }_{{{k}_{1}}}}}}{\partial {{\psi }_{{{k}_{2}}}}} \right\} \\ 
		& =\text{Tr}\left\{ \frac{\partial {{\boldsymbol{G}}^{-1}}}{\partial {{\psi }_{{{k}_{2}}}}}\frac{\partial \boldsymbol{G}}{\partial {{\theta }_{{{k}_{1}}}}} \right\}+\text{Tr}\left\{ {{\boldsymbol{G}}^{-1}}\frac{\partial \boldsymbol{G}}{\partial {{\theta }_{{{k}_{1}}}}\partial {{\psi }_{k2}}} \right\},
	\end{split}
\end{align}
\begin{align}
	\begin{split}
		\mathcal{E}\left\{ \frac{\partial {{{\boldsymbol{y}}}^{\text{H}}}{{\boldsymbol{G}}^{-1}}\boldsymbol{y})}{\partial {{\theta }_{{{k}_{1}}}}\partial {{\psi }_{{{k}_{2}}}}} \right\}=\text{Tr}\left\{ \frac{\partial {{\boldsymbol{G}}^{-1}}}{\partial {{\theta }_{{{k}_{1}}}}\partial {{\psi }_{{{k}_{2}}}}}\boldsymbol{G} \right\},
	\end{split}
\end{align}
therefore, we can get
\begin{align}
	\begin{split}
		\boldsymbol{F}_{{{k}_{1}},{{k}_{2}}}^{1,2}=\text{Tr}\left\{ \frac{\partial \boldsymbol{G}}{\partial {{\theta }_{{{k}_{1}}}}}{{\boldsymbol{G}}^{-1}}\frac{\partial \boldsymbol{G}}{\partial {{\psi }_{{{k}_{2}}}}}{{\boldsymbol{G}}^{-1}} \right\}.
	\end{split}
\end{align}
Next we have
\begin{align}
	\begin{split}
		{{\boldsymbol{F}}_{2,1}}=-\mathcal{E}\left\{ \left. \frac{\partial \ln f(\boldsymbol{y};\boldsymbol{\theta },\boldsymbol{\psi })}{\partial \boldsymbol{\psi }\partial \boldsymbol{\theta }} \right|\boldsymbol{\theta },\boldsymbol{\psi } \right\},
	\end{split}
\end{align}
the ${{k}_{1}}$,  ${{k}_{2}}$-th entry of Fisher information matrix ${{\boldsymbol{F}}_{2,1}}$ can be obtained as
\begin{align}
	\begin{split}
		& \boldsymbol{F}_{{{k}_{1}},{{k}_{2}}}^{2,1}=\frac{\partial \ln \det \left\{ \boldsymbol{G} \right\}}{\partial {{\psi }_{{{k}_{1}}}}\partial {{\theta }_{{{k}_{2}}}}}+\mathcal{E}\left\{ \frac{\partial {{{\boldsymbol{y}}}^{\text{H}}}{{\boldsymbol{G}}^{-1}}\boldsymbol{y})}{\partial {{\psi }_{{{k}_{1}}}}\partial {{\theta }_{{{k}_{2}}}}} \right\} \\ 
		& =\text{Tr}\left\{ \frac{\partial \boldsymbol{G}}{\partial {{\psi }_{{{k}_{1}}}}}{{\boldsymbol{G}}^{-1}}\frac{\partial \boldsymbol{G}}{\partial {{\theta }_{{{k}_{2}}}}}{{\boldsymbol{G}}^{-1}} \right\}.
	\end{split}
\end{align}
Then we can get
\begin{align}
	\begin{split}
		{{\boldsymbol{F}}_{2,2}}=-\mathcal{E}\left\{ \left. \frac{\partial \ln f(\boldsymbol{y};\boldsymbol{\theta },\boldsymbol{\psi })}{\partial \boldsymbol{\psi }\partial \boldsymbol{\psi }} \right|\boldsymbol{\theta },\boldsymbol{\psi } \right\},
	\end{split}
\end{align}
and the ${{k}_{1}}$,  ${{k}_{2}}$-th entry of Fisher information matrix ${{\boldsymbol{F}}_{2,2}}$ can be obtained as
\begin{align}
	\begin{split}
		& \boldsymbol{F}_{{{k}_{1}},{{k}_{2}}}^{2,2}=\frac{\partial \ln \det \left\{ \boldsymbol{G} \right\}}{\partial {{\psi }_{{{k}_{1}}}}\partial {{\psi }_{{{k}_{2}}}}}+\mathcal{E}\left\{ \frac{\partial {{{\boldsymbol{y}}}^{\text{H}}}{{\boldsymbol{G}}^{-1}}\boldsymbol{y})}{\partial {{\psi }_{{{k}_{1}}}}\partial {{\psi }_{{{k}_{2}}}}} \right\} \\ 
		& =\text{Tr}\left\{ {{\boldsymbol{G}}^{-1}}\frac{\partial \boldsymbol{G}}{\partial {{\psi }_{{{k}_{1}}}}}{{\boldsymbol{G}}^{-1}}\frac{\partial \boldsymbol{G}}{\partial {{\psi }_{{{k}_{2}}}}} \right\}.
	\end{split}
\end{align}
Finally, the CRLB of the ${k}$-th DOA estimation can be got from Fisher information matrix 
\begin{align}
	\begin{split}
		\operatorname{var}\left\{ {{\theta }_{k}} \right\}\ge \sum\limits_{k=0}^{K-1}{{{\left[ {{\boldsymbol{F}}^{-1}} \right]}_{k,k}}}.
	\end{split}
\end{align} 
The inverse of the Fisher information matrix is hard to be obtained. Usually, a lower bound of FIM is used to describe the DOA estimation performance \cite{ref50}.
\begin{align}
	\begin{split}
		\operatorname{var}\left\{ {{\theta }_{k}} \right\}\ge \sum\limits_{k=0}^{K-1}{\boldsymbol{F}_{k,k}^{-1}}.
	\end{split}
\end{align}

\section{Simulation Results}

In this section, the simulation results are given to show the performance of the proposed NC-ANM method in LPDF system. All the simulation results are obtained on a PC with Matlab R2018b with a 2.6 GHz Intel Core i7 and 16 GB of RAM. The number of Monte Carlo simulations is $M=200$, and the other simulation parameters are given in Table~\ref{tab:table1}. The IRS with $32$ elements is controlled by a FPGA to realize the phase responses of $\left\{ 0,\pi  \right\}$. Let the amplitude ${{A}_{0,p}}={{A}_{1,p}}=\cdots ={{A}_{N-1,p}}=1$, then the entry of ${\boldsymbol{e}(p)}$ is randomly chosen as $1$ or $-1$. 

\begin{table} [!t]     
	\caption{Simulation Parameters} \label{tab:table1}   
	\centering
	\begin{tabular}{cc}
		\toprule[1pt] 
		Parameter &Value \\
		\midrule[1pt]     
		The number of IRS $N$ & $32$\\
		The number of signals $K$ & $3$\\
		The number of measurements $P$ & $32$\\
		The sparsity of signal $S$ & $300$ \\
		The space between elements $d$ & $0.5\lambda$\\
		The DOA range of direction & $[-{50}^{\text{o}},{50}^{\text{o}}]$\\
		The SNR of received signal & $20$ dB\label{tab:table 1}\\
		\bottomrule[1pt] 
	\end{tabular}
\end{table}
\begin{figure}[!t]	
	\centering
	\includegraphics[width=1\columnwidth]{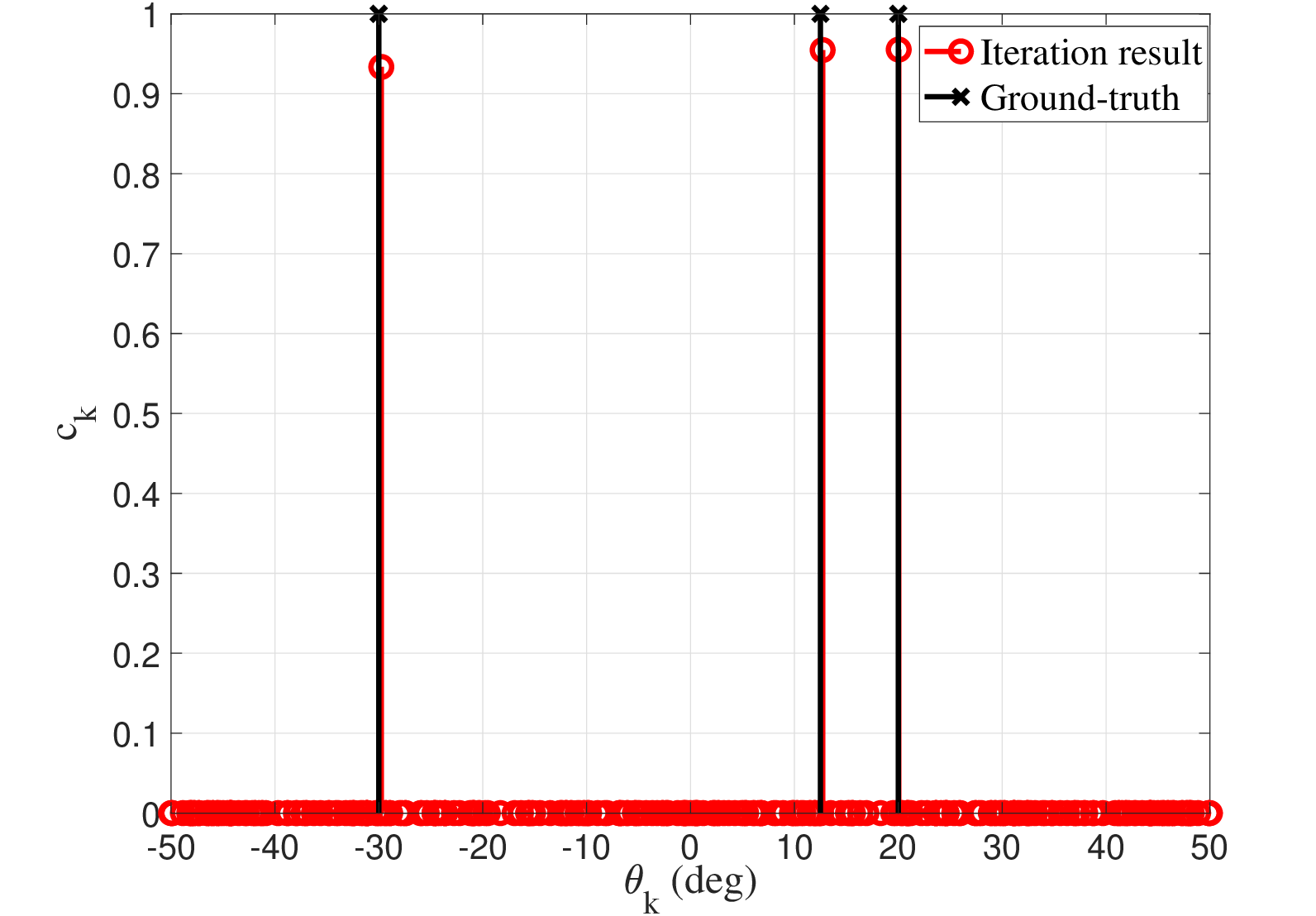}
	\caption{The sparse reconstruction result.}
	\label{fig.iteration}
\end{figure}

First, with the simulation parameters in Table~\ref{tab:table1}, we try to estimate $3$ signals locating at the directions of $-30.01^{\text{o}}$, $12.51^{\text{o}}$ and  $20.00^{\text{o}}$. The sparsity $S=300$, the threshold $T$ is set to the $\frac{S}{2}$-th value of $\boldsymbol{{c}}^{q}$ after sorting, and the measurements $P=32$. When the proposed method NC-ANM is adopted, the sparse reconstruction result shown in Fig.~\ref{fig.iteration} is $-29.76^{\text{o}}$, $12.72^{\text{o}}$ and $20.04^{\text{o}}$. It shows that the proposed method can reconstruct signals effectively.

\begin{figure} [!t]	
	\centering
	\includegraphics[width=1\columnwidth]{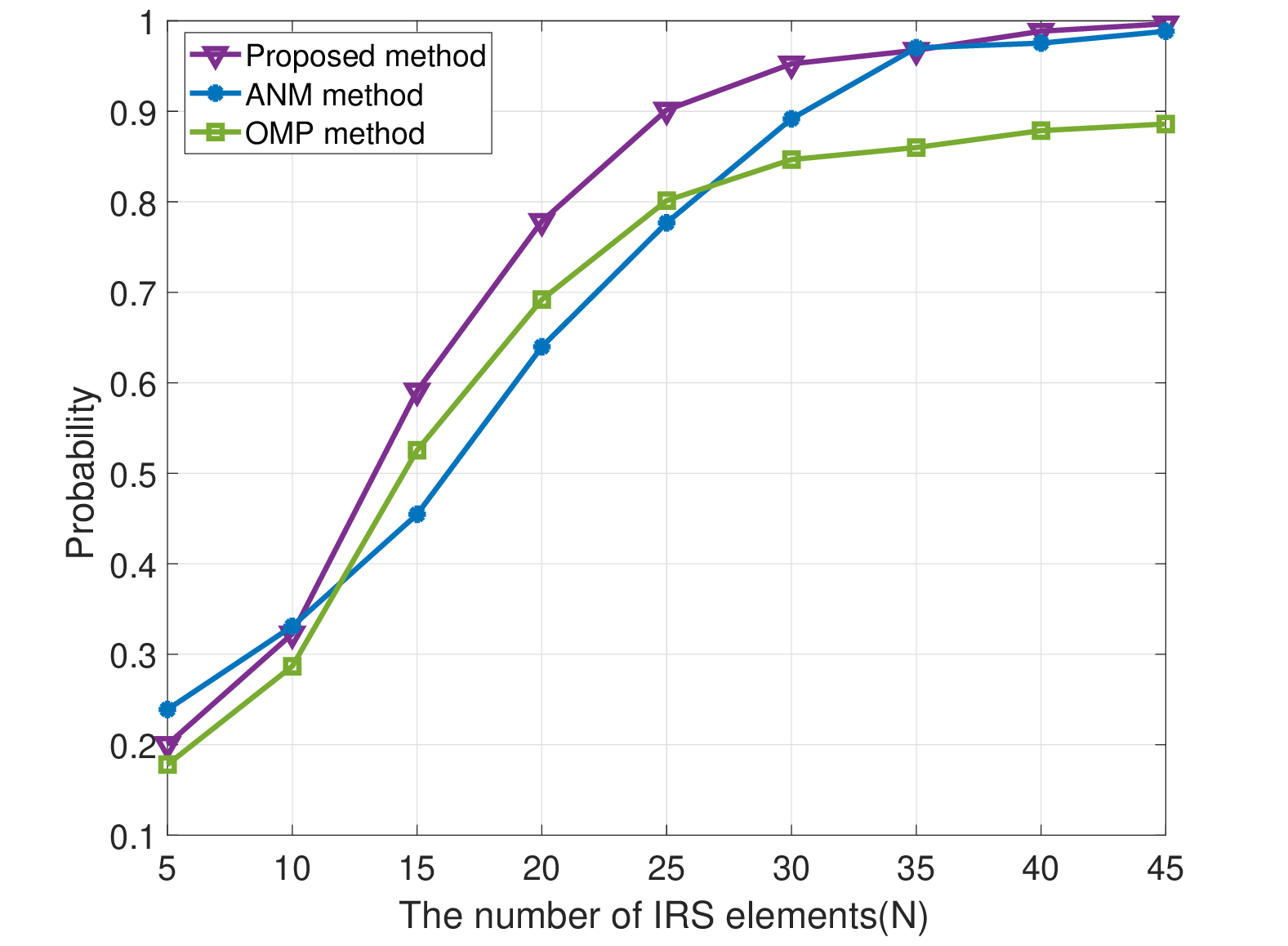}
	\caption{The probability for signal reconstruction with different IRS element numbers.}
	\label{fig.unit_pro}
\end{figure}
\begin{figure}
	\centering
	\includegraphics[width=1\columnwidth]{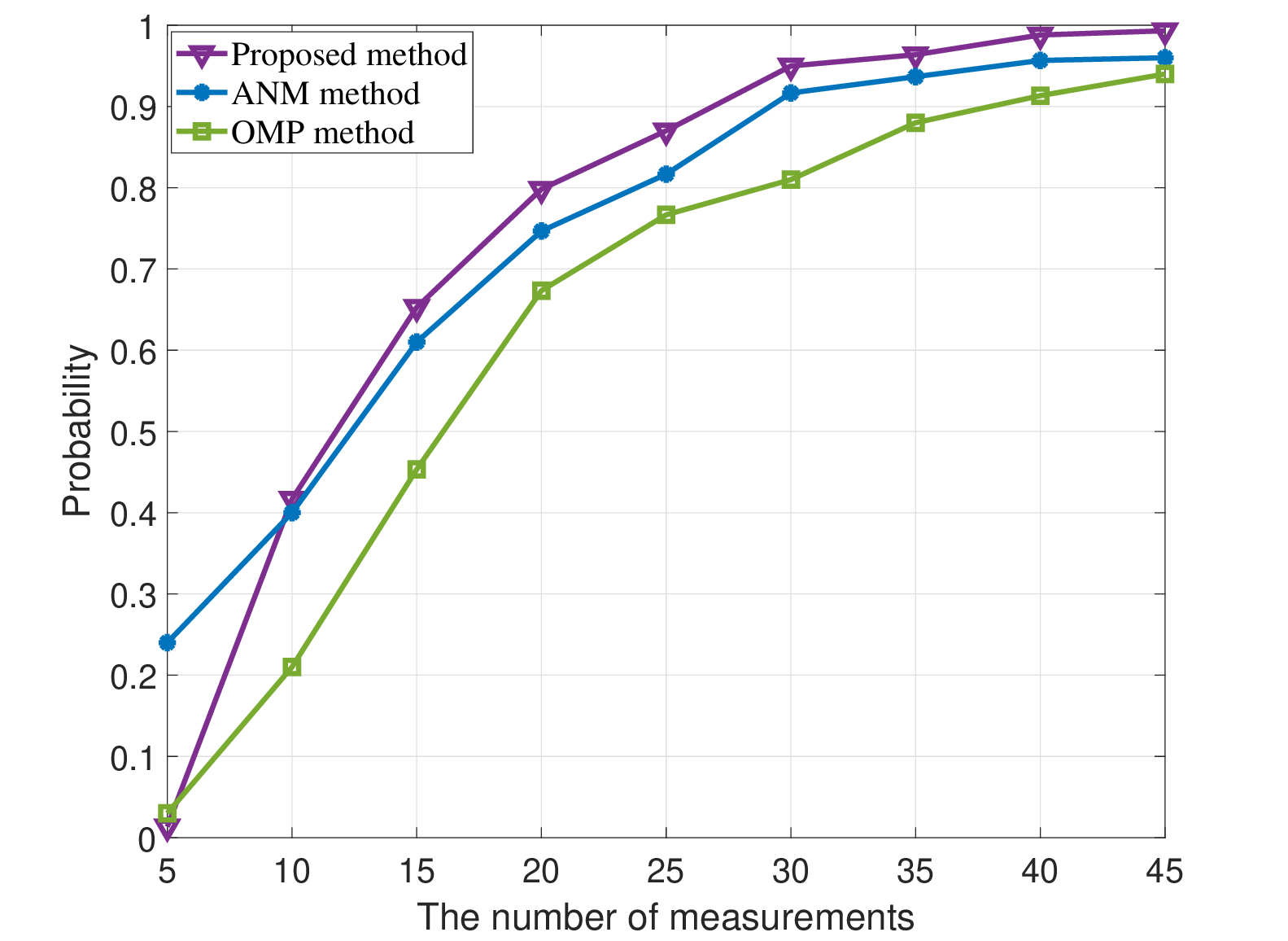}
	\caption{The probability for signal reconstruction with different IRS element numbers.}
	\label{fig.measurement_pro}
\end{figure}

Then the reconstruction probabilities of the proposed method, ANM and OMP are shown in Fig.~\ref{fig.unit_pro} and Fig.~\ref{fig.measurement_pro} with different numbers of IRS elements and measurements. When $S=300$, $P=32$, and $N$ increases from $10$ to $50$ with the step of $5$, the probabilities of the three methods are all improved and the proposed method performs best. If the size of IRS is large enough, the probability can approach $1$. Similarily, the probabilities of these methods are also improved with measurements number increasing. The proposed method outperforms the ANM and OMP when $P<10$.

\begin{figure}[!t]	
	\centering
	\includegraphics[width=1\columnwidth]{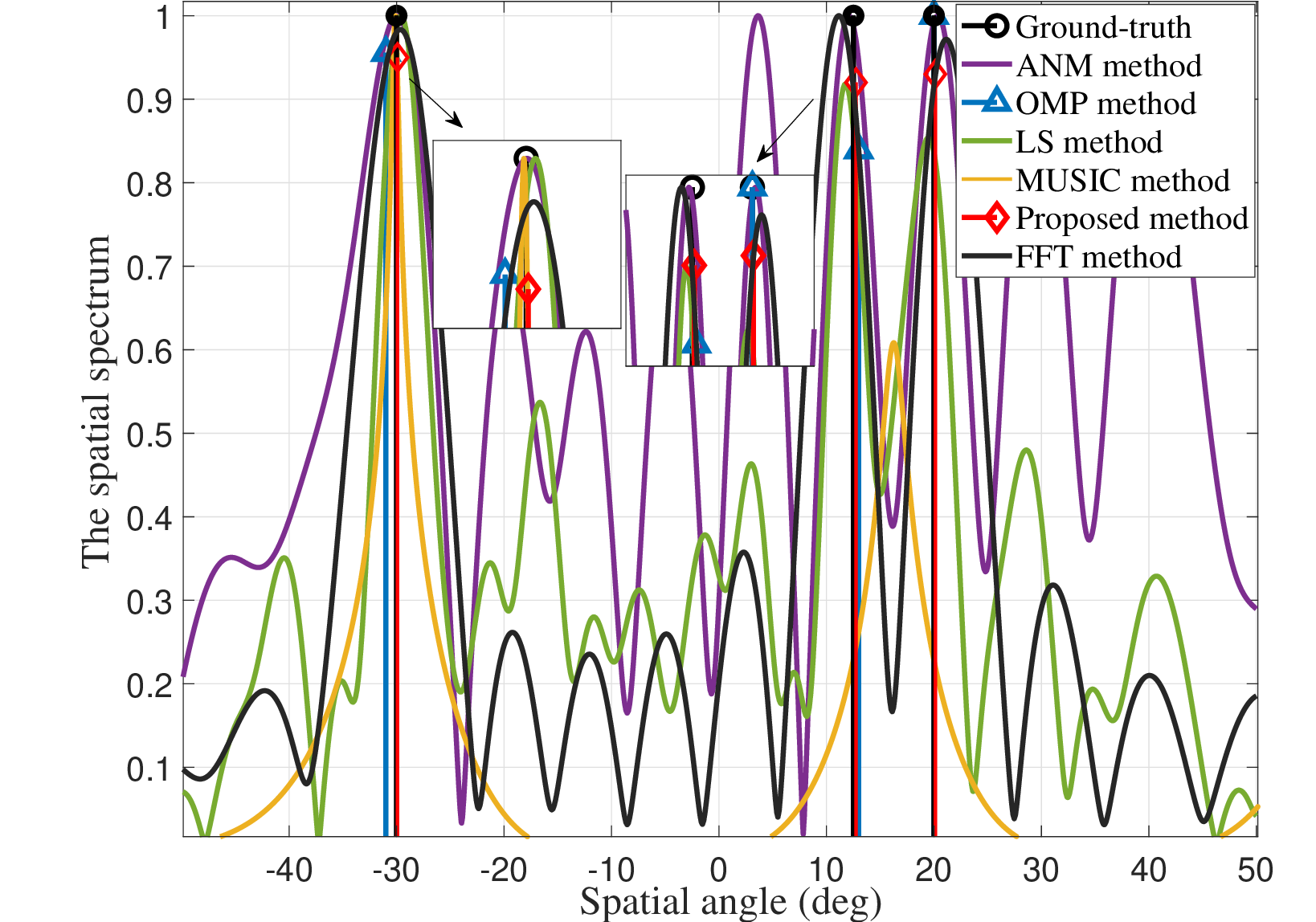}
	\caption{The spatial spectrum for DOA estimation.}
	\label{fig.methods}
\end{figure}

Additionally, we use the proposed algorithm NC-ANM for DOA estimation, and compare its performance with the traditional methods, including ANM, OMP, LS, fast Fourier transform (FFT)\cite{ref51} and MUSIC. The root-mean-square error (RMSE) of each method is calculated to measure the performance of DOA estimation. RMSE is defined as
\begin{align}
	\begin{split}
		\text{RMSE}\triangleq \sqrt{\frac{1}{K}\sum\nolimits_{m=0}^{M-1}{\sum\nolimits_{k=0}^{K-1}{{{({{\theta }_{m,k}}-{{{\hat{\theta }}}_{m,k}})}^{2}}}}}.
	\end{split}
\end{align}
where ${{\theta }_{m,k}}$ denotes the ground-truth DOA of $k$-th signal during $m$-th simulations, $m=0,1,...,M$, $k=0,1,...,K$. ${{\hat{\theta }}_{M,k}}$ denotes the corresponding estimated result. When the proposed method, ANM, LS, MUSIC, FFT and OMP are adopted, the simulation results are shown in Fig.~\ref{fig.methods} and the  RMSEs are shown in Table~\ref{tab:table2}.

\begin{table} 
	\caption{Estimation Results of DOA} \label{tab:table2}   
	\centering
	\begin{tabular}{ccccc}
		\toprule[1pt] 
		Methods &Signal 1 &Signal 2 &Signal 3 &RMSE\\
		&(deg)&(deg)&(deg)&(deg)\\
		\midrule[1pt]     
		Ground-truth &$-30.01$ & $12.51$ & $20.00$ & $-$\\
		ANM &$-29.97$ & $12.07$ & $20.37$ & $0.44$\\
		LS &$-29.58$ & $11.77$ & $19.48$ & $0.58$\\
		OMP &$-30.50$ & $12.50$ & $20.50$ & $0.48$\\
		FFT &$-29.68$ & $12.38$ & $21.20$ & $0.80$\\
		MUSIC &$-30.12$ & $-6.56$ & $16.26$ & $2.55$\\
		Proposed method &$-29.76$ & $12.72$ & $20.04$ & $0.19$\\
		\bottomrule[1pt] 
	\end{tabular}
\end{table}

The proposed method is also compared with the traditional ANM method. The DOA estimations using the proposed NC-ANM method are $-29.76^{\text{o}}$, $12.72^{\text{o}}$ and $20.04^{\text{o}}$, where the RMSE is $0.19^{\text{o}}$. Under the same condition, the estimated DOAs using traditional ANM are  $-29.97^{\text{o}}$, $12.07^{\text{o}}$, $20.37^{\text{o}}$, and the RMSE is $0.44^{\text{o}}$. It is ${56.3\%}$ worse than the proposed method. Besides that, the traditional ANM adopts CVX toolbox for solving the SDP problem, which has high complexity and derives the suboptimum solution because of the relaxation. In the OMP method, the spatial angle is discretized into grids with the step of $0.5^{\text{o}}$, where the ground-truth DOAs are not on their grid exactly. Compared with the proposed method and ANM, the OMP has a higher RMSE of $0.48^{\text{o}}$ due to the grid mismatch.

\begin{table} 
	\caption{Computation Time} \label{tab:table3}   
	\centering
	\begin{tabular}{cccccc}
		\toprule[1pt] 
		~ &ANM &LS &OMP &FFT&Proposed method\\
		\midrule[1pt]     
		Time(s) &$1.99$ & $0.35$ & $0.46$ & $0.80$ & $1.06$	\\	
		\bottomrule[1pt] 
	\end{tabular}
\end{table}

\begin{figure}[!t]
	\centering
	\includegraphics[width=1\columnwidth]{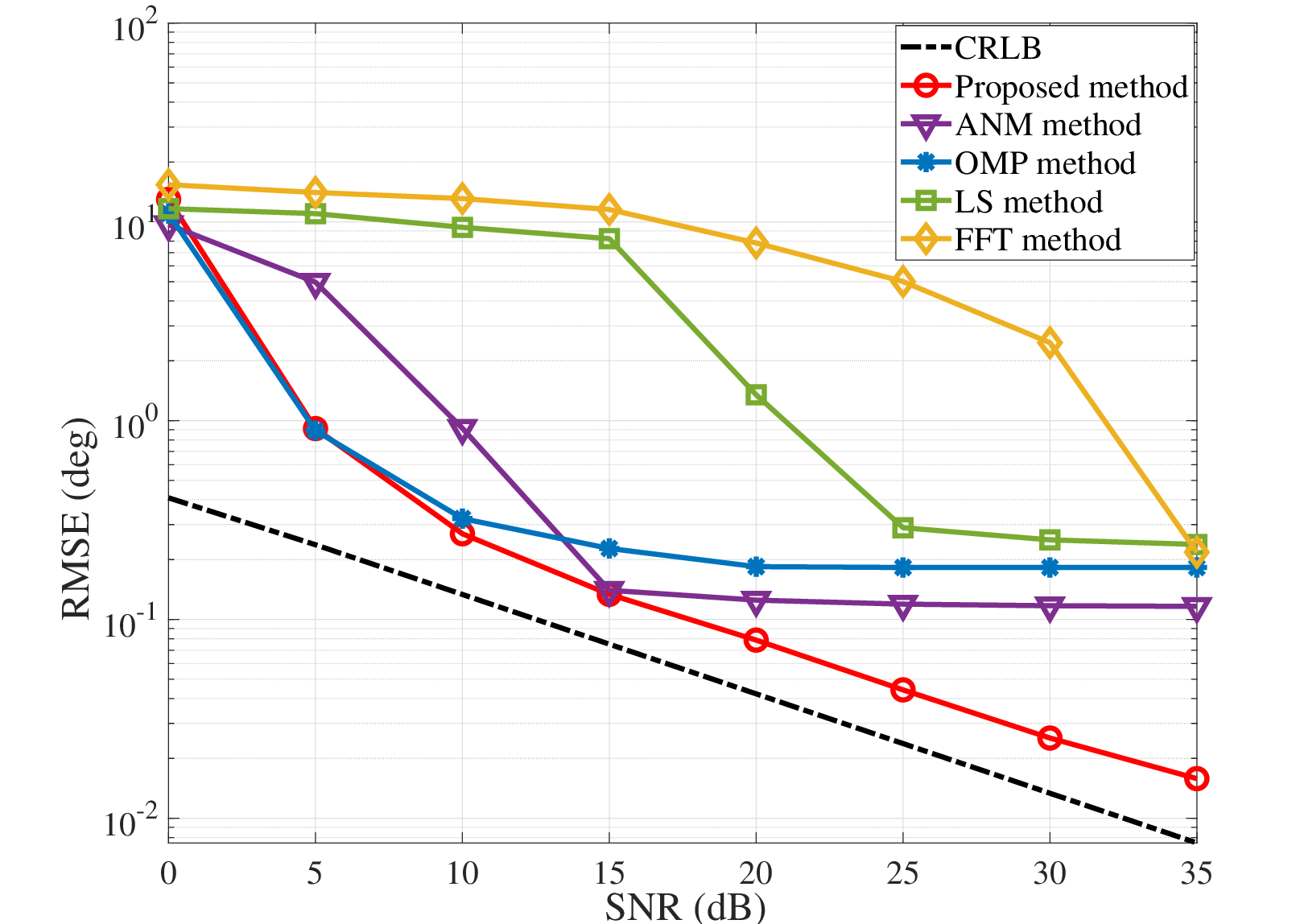}
	\caption{The performance of DOA estimation with different SNRs.}
	\label{RMSE_snr}
\end{figure}

\begin{figure}[!t]
	\centering
	\includegraphics[width=1\columnwidth]{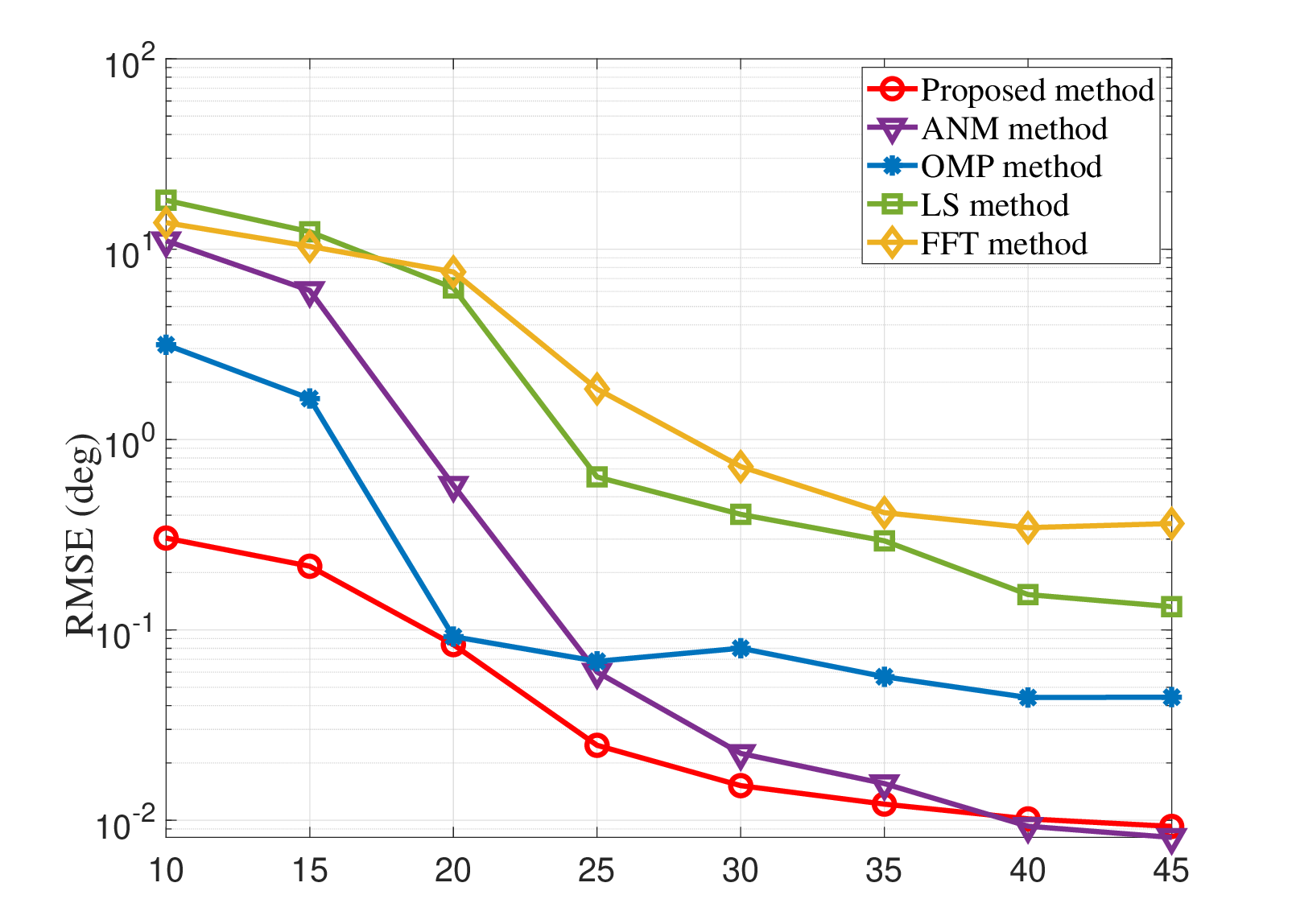}
	\caption{The performance of DOA estimation with different numbers of IRS.}
	\label{fig. RMSE_units}
\end{figure}

Different from the sparse-based methods, the LS method, FFT method and MUSIC method have worse performance. 
The estimated results of LS are  $-29.58^{\text{o}}$, $11.77^{\text{o}}$, $19.48^{\text{o}}$ with RMSE of $0.58^{\text{o}}$. It is $66.9\%$ higher than the proposed method because it has an unstable solution due to matrix inversion. The MUSIC method applied in the LPDF system fails because of single measurement. Even though considering the reference~\cite{ref52} and using the Single-snapshot MUSIC method under single-snapshot data, we can only get one inaccurate DOA at $-30.12^{\text{o}}$. Its performance has been greatly weakened compared with its in the phased array because of the limitation of measurement. Similarly, although FFT can obtain the DOA estimation of $-29.68^{\text{o}}$, $12.38^{\text{o}}$, $21.20^{\text{o}}$, its RMSE $0.80^{\text{o}}$ is the highest.

The proposed method can achieve the best performance with relative higher computational complexity of $10.44$s. If we reduce the iteration times and sparsity rate, the RMSE reduces to $0.43^{\text{o}}$ and the computational time of each method is shown in Table~\ref{tab:table3}. The proposed method has lower complexity than ANM with similar RMSE.

\begin{figure}[!t]
	\centering
	\includegraphics[width=1\columnwidth]{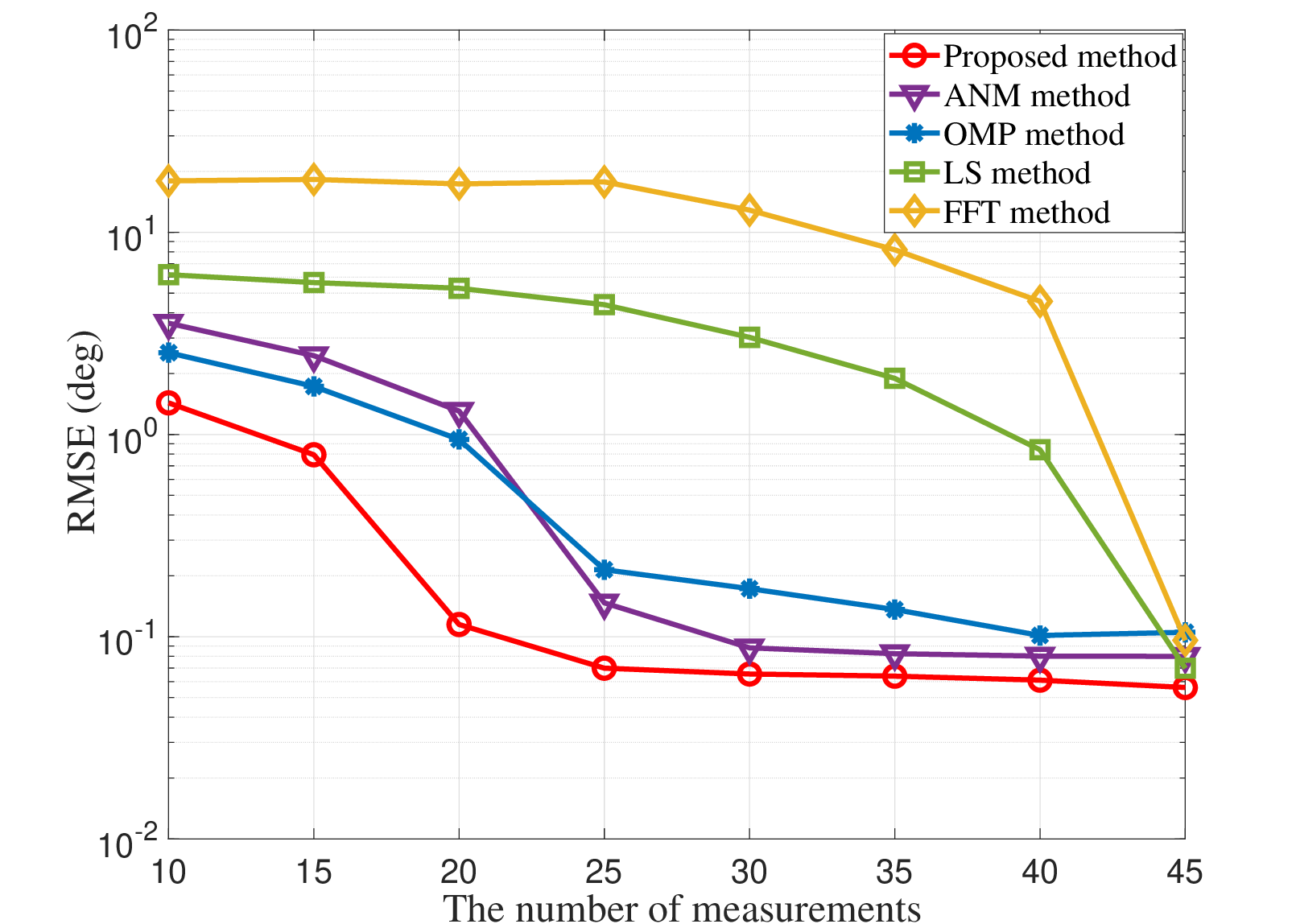}
	\caption{The performance of DOA estimation with different measurements.}
	\label{fig. RMSE_measurement}
\end{figure}

Then, the performance versus SNR with different methods is studied, where the elements number of IRS $N=32$, the measurement number $P=32$, and the SNR is varied from $0$ dB to $35$ dB with the step of $5$ dB. With the MUSIC algorithm invalid, we no longer consider its performance. The simulation results are shown in Fig.~\ref{RMSE_snr}. It can be observed that the RMSEs of the sparse-based methods are much better than FFT and LS. When the SNR is lower than $10$ dB, the estimation performances of the proposed method and OMP are almost the same. With the increase of SNR, the performance of these methods gradually improves. Until the SNR achieves $15$dB, the proposed method has the best performance among these methods.

For different sizes of IRS, the DOA estimation performance is shown in Fig.~\ref{fig. RMSE_units}, where the measurement number $P=32$, $\text{SNR}=20$ dB and the number of IRS elements $N$ varies from $10$ to $45$ with the step of $5$. It can be observed that the performance of these methods is improved with more IRS elements and the proposed method has the lowest RMSE until $N=40$. When elements are more than $40$, the estimation performance has a platform errors.  

Additionally, when $\text{SNR}=20$ dB, the IRS elements number $N=32$, the measurements number $P$ is varied from $10$ to $45$ with step size $5$, the corresponding estimation performance of these methods versus IRS elements is given in Fig.~\ref{fig. RMSE_measurement}. The RMSEs of the proposed method, ANM and OMP decrease significantly before $P=25$. LS and FFT methods can achieve the same performance when the number of measurements is large enough. Moreover, the proposed method also has the best DOA estimation performance among these methods.

Through the above simulations, we can conclude that the proposed method has better performance via overcoming the weekness of existing methods. The ANM method has high complexity and derives the suboptimum solution. The OMP method has the limitation of  grid mismatch. The LS method is unstable due to the matrix inversion. The FFT method performs badly with less snapshots or measurements.

\section{Conclusion}
In this paper, the model of LPDF system and received signal have been formulated, and the DOA estimation problem has been considered. In the LPDF system, IRS has been used to receive the signals in the blind spots in the advantage of a low cost. For the DOA estimation problem, the NC-ANM method has been proposed to achieve high accuracy and low complexity. The NC-ANM method has solved the non-convex optimization problem directly via gradient threshold iteration, where the random perturbation is added to avoid the local optimum solution. Simulation results demonstrate that the proposed method outperforms the existing methods in the scenario with LPDF system. Future work will focus on the proposed DOA estimation method with gain-phase errors.

\begin{appendices}

    \section{The Proof of the Proposition}\label{appA}
According to the Definition~{\ref{Def1}}, the formula obtained by exchanging $x$ and $y$ is
\begin{align}
	\begin{split}
		\nabla & f(x)^{\text{T}}(y-x)+\frac{l}{2}\left\|{y-x}\right\|^2_2 \\ & \leq f(y)-f(x) \\ & \leq \nabla f(x)^{\text{T}}(y-x)+\frac{L}{2}\left\|{y-x}\right\|^2_2.
	\end{split}
\end{align}
Add it with ({\ref{eq35}})
\begin{align}
	\begin{split}
		(\nabla f(x)-\nabla f(y))^{\text{T}}(x-y) \leq L \left\|{x-y}\right\|^2_2.
	\end{split}
\end{align}
According to Cauchy inequality 
\begin{align}
	\begin{split}
		& (\nabla f(x)-\nabla f(y))^{\text{T}}(x-y)\left[(x-y)^{\text{T}}(\nabla f(x)-\nabla f(y)) \right] \\ = & \left\| (\nabla f(x)-\nabla f(y))^{\text{T}}(x-y) \right\|^2_2 \\ \leq & \left\| \nabla f(x)-\nabla f(y) \right\|^2_2\left\| x-y \right\|^2_2,
	\end{split}
\end{align}
we can get
\begin{align}
	\begin{split}
		& \left[(x-y)^{\text{T}}(\nabla f(x)-\nabla f(y)) \right] \\ \leq & 
		\left\| (\nabla f(x)-\nabla f(y))\right\|_2 \left\| ( x-y)\right\|_2\\ \leq & L\left\| x-y \right\|^2_2, 
	\end{split}
\end{align}
\begin{align}
	\begin{split} 
		\left\| (\nabla f(x)-\nabla f(y))\right\|_2 \leq L\left\| x-y \right\|_2.
	\end{split}
\end{align}

Then the following inequality can be obtained
\begin{align}
	\begin{split} 
		& \left\|x-y-\zeta \nabla f(x)+\zeta \nabla f(y) \right\|^2_2 \\ = & \left\| x-y \right\|_2^2-2\zeta(x-y)^{\text{T}}(\nabla f(x)-\nabla f(y)) \\ & + \zeta ^2 \left\| (\nabla f(x)-\nabla f(y))\right\|_2^2 \\ \leq & (1-2\zeta l+\zeta^2 L^2)\left\| x-y \right\|_2^2 \\ =& \rho^2\left\| x-y \right\|_2^2,
	\end{split}
\end{align}
where $\rho= \sqrt{1-2\zeta l + \zeta^2L^2}$. Thus, we can get
\begin{align}
	\begin{split} 
		\left\|x-y-\zeta \nabla f(x)+\zeta \nabla f(y) \right\|_2 \leq  \rho \left\| x-y \right\|_2.
	\end{split}
\end{align}
From
\begin{align}
	\begin{split} 
		& \left\|x-y \right\|_2-\zeta \left\|\nabla f(x)- \nabla f(y) \right\|_2 \\ \leq & \left\|x-y-\zeta \nabla f(x)+\zeta \nabla f(y) \right\|_2 \\ \leq & \rho \left\| x-y \right\|_2,
	\end{split}
\end{align}
\begin{align}
	\begin{split} 
		& \zeta \left\|\nabla f(x)- \nabla f(y) \right\|_2 -\left\|x-y \right\|_2 \\ \leq & \left\|x-y-\zeta \nabla f(x)+\zeta \nabla f(y) \right\|_2 \\ \leq & \rho \left\| x-y \right\|_2,
	\end{split}
\end{align}
we can get
\begin{align}
	\begin{split} 
	 \frac{1-\rho}{\zeta}\left\|x-y \right\|_2 \leq & \left\|\nabla f(x)-\nabla f(y) \right\|_2 \\ \leq &  \frac{1+\rho}{\zeta} \left\| x-y \right\|_2.
	\end{split}
\end{align}
Finally, the inequality~({\ref{eq36}}) can be obtained
\begin{align}
	\begin{split} 
		& f(x)-f(y)- \langle {\nabla f(y),x-y} \rangle \\ = & \int\limits_0^1 {\langle{\nabla f(y+\delta (x-y))-\nabla f(y),x-y} \rangle {\text{d} \delta}} \\ \leq & \int\limits_0^1 {\left\| \nabla {f(y+\delta (x-y))-\nabla f(y)}\right\|_2 \left\|{x-y}\right\|_2{\text{d} \delta}} \\ \leq & \int\limits_0^1 {\frac{1+\rho}{2\zeta}\left\|x-y\right\|^2_2} \delta {\text{d} \delta} \\ \leq & \frac{1+\rho}{2\zeta}\left\|x-y\right\|^2_2.
	\end{split}
\end{align}

\end{appendices}


\newpage

\balance
\vspace{11pt}
\vspace{-45pt}
\begin{IEEEbiography}
[{\includegraphics[width=1in,height=1.25in,clip,keepaspectratio]{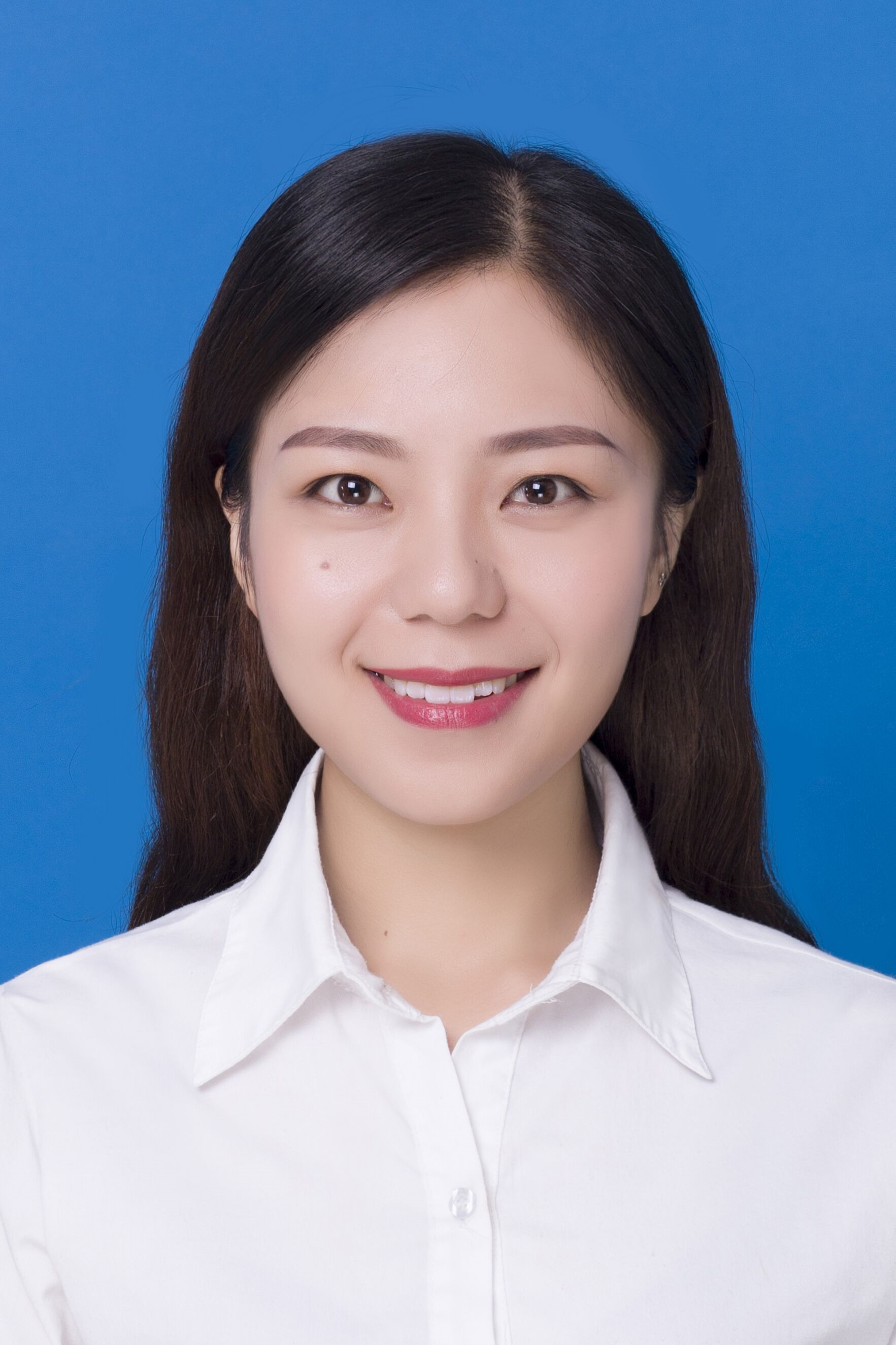}}]{Yangying Zhao}
(Member, IEEE) was born in Heilongjiang, China, in 1993. She received the B.E. and M.S. degrees from the School of Electronic and Optical Engineering, Nanjing University of Science and Technology, China, in 2015 and 2019. She is currently pursuing the Ph.D. degree with the State Key Laboratory of Millimeter Waves, Southeast University, Nanjing, China.

Her research interests include radar signal processing and millimeter wave communication.
\end{IEEEbiography}
\vspace{-33pt}
\begin{IEEEbiography}
	[{\includegraphics[width=1in,height=1.25in,clip,keepaspectratio]{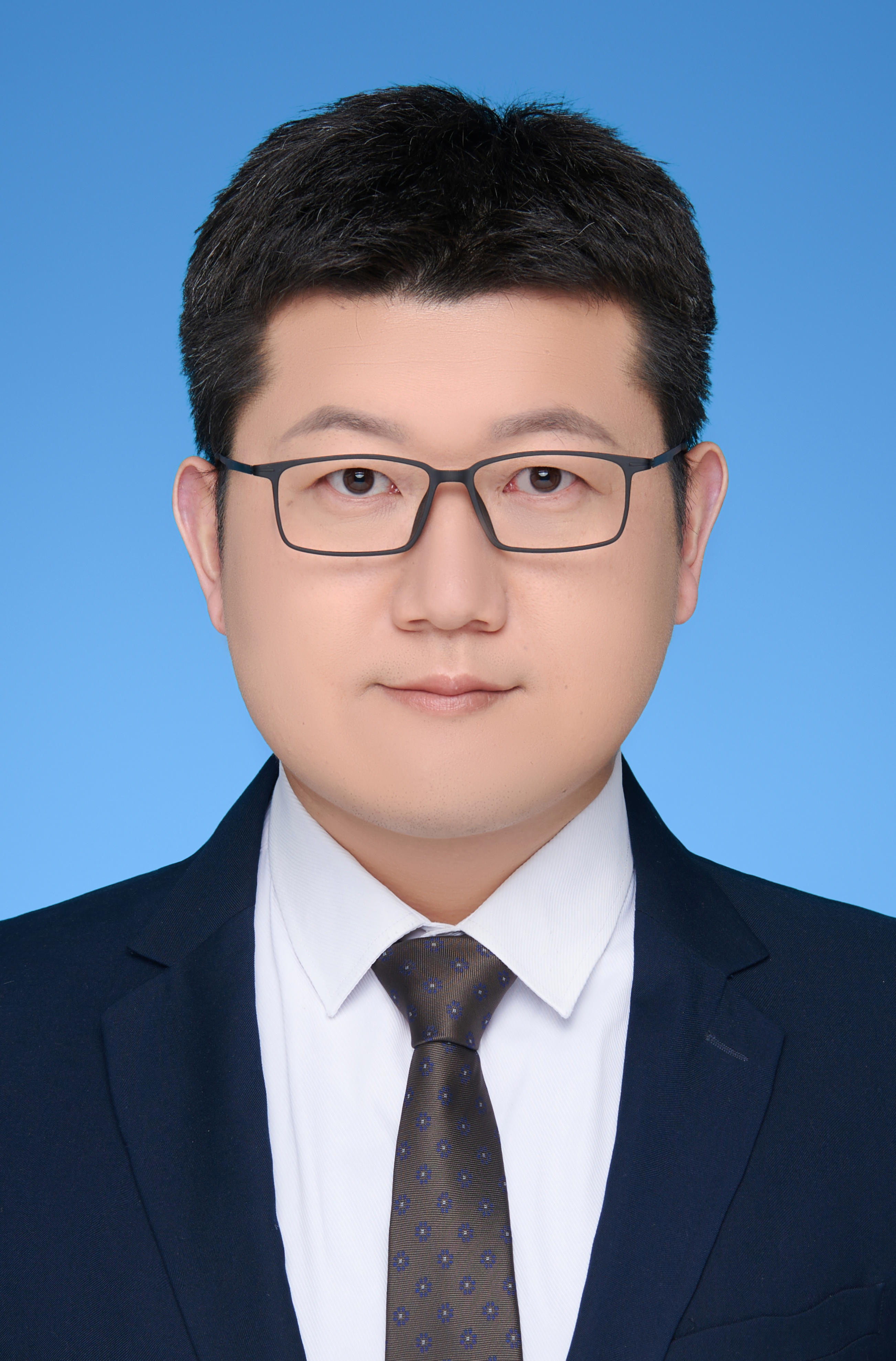}}]{Peng Chen} (Senior Member, IEEE) was born in Jiangsu, China, in 1989. He received the B.E. and Ph.D. degrees from the School of Information Science and Engineering, Southeast University, Nanjing, China, in 2011 and 2017 respectively. From March 2015 to April 2016, he was a Visiting Scholar with the Department of Electrical Engineering, Columbia University, New York, NY, USA. He is currently an Associate Professor with the State Key Laboratory of Millimeter Waves, Southeast University. His research interests include target localization, super-resolution reconstruction, and array signal processing. 
	
	He is a Jiangsu Province Outstanding Young Scientist. He has served as an IEEE ICCC Session Chair, and won the Best Presentation Award in 2022 (IEEE ICCC). He was invited as a keynote speaker at the IEEE ICET in 2022. He was recognized as an exemplary reviewer for IEEE WCL in 2021, and won the Best Paper Award at IEEE ICCCCEE in 2017.

\end{IEEEbiography}
\vspace{-25pt}
\begin{IEEEbiography}
	[{\includegraphics[width=1in,height=1.25in,clip,keepaspectratio]{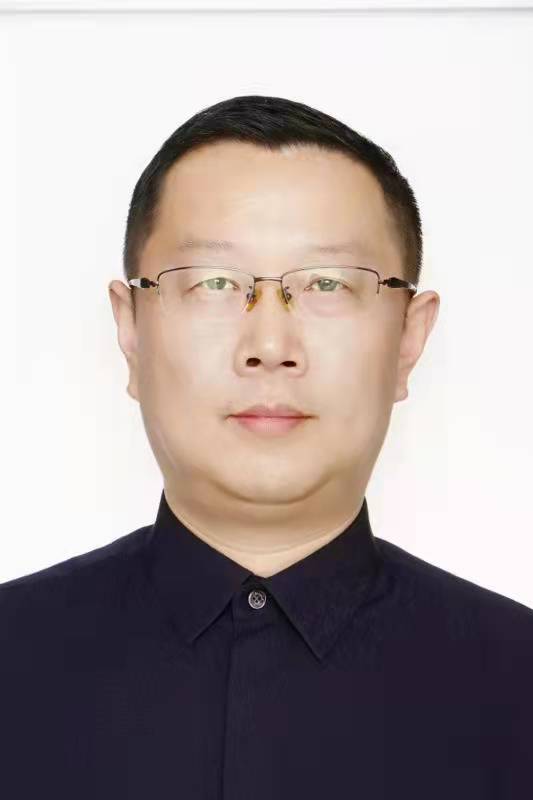}}]{Zhenxin Cao}
	(Member, IEEE) was born in May 1976. He received the M.S. degree from Nanjing University of Aeronautics and Astronautics, Nanjing, China, in 2002 and the Ph.D. degree from the School of Information Science and Engineering, Southeast University, Nanjing, China, in 2005. From 2012 to 2013, he was a Visiting Scholar with North Carolina State University Since 2005, he has been with the State Key Laboratory of Millimeter Waves, Southeast University, where he is currently a professor.
	
	His research interests include antenna theory and application.
\end{IEEEbiography}
\vspace{-25pt}
\begin{IEEEbiography}
	[{\includegraphics[width=1in,height=1.25in,clip,keepaspectratio]{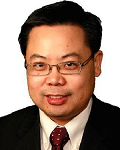}}]{Xianbin Wang}
	(Fellow, IEEE) received the Ph.D. degree in electrical and computer engineering from National University of Singapore, in 2001. He is a Professor and Tier-I Canada Research Chair at Western University, Canada. Prior to joining Western University, he was with Communications Research Centre (CRC) Canada as a Research Scientist/Senior Research Scientist between July 2002 and December 2007. From January 2001 to July 2002, he was a System Designer with STMicroelectronics, where he was responsible for the system design of DSL and Gigabit Ethernet chipsets. His current research interests include 5G technologies, Internet of Things, communications security, machine learning and locationing technologies. Dr. Wang has over 300 peer-reviewed journal and conference papers, in addition to 26 granted and pending patents and several standard contributions. Dr. Wang is a Fellow of Canadian Academy of Engineering and an IEEE Distinguished Lecturer. He has received many awards and recognitions, including Canada Research Chair, CRC Presidents Excellence Award, Canadian Federal Government Public Service Award, Ontario Early Researcher Award and five IEEE Best Paper Awards. He currently serves as an Editor/Associate Editor for IEEE Transactions on Communications, IEEE Transactions on Broadcasting, and IEEE Transactions on Vehicular Technology and he was also an Associate Editor for IEEE Transactions on Wireless Communications between 2007 and 2011, and IEEE Wireless Communications Letters between 2011 and 2016. Dr. Wang was involved in many IEEE conferences including GLOBECOM, ICC, VTC, PIMRC, WCNC, and CWIT, in different roles such as symposium chair, tutorial instructor, track chair, session chair and TPC co-chair.
\end{IEEEbiography}

\vfill
\end{document}